\documentstyle[epsfig,psfig]{mn}     
     
\newif\ifAMStwofonts     
     
\ifoldfss     
  \ifCUPmtlplainloaded \else     
    \NewTextAlphabet{textbfit} {cmbxti10} {}     
    \NewTextAlphabet{textbfss} {cmssbx10} {}     
    \NewMathAlphabet{mathbfit} {cmbxti10} {} 
    \NewMathAlphabet{mathbfss} {cmssbx10} {} 
  \fi     
  \ifAMStwofonts     
    \ifCUPmtlplainloaded \else     
      \NewSymbolFont{upmath} {eurm10}     
      \NewSymbolFont{AMSa} {msam10}     
      \NewMathSymbol{\upi}     {0}{upmath}{19}     
      \NewMathSymbol{\umu}     {0}{upmath}{16}     
      \NewMathSymbol{\upartial}{0}{upmath}{40}     
      \NewMathSymbol{\leqslant}{3}{AMSa}{36}     
      \NewMathSymbol{\geqslant}{3}{AMSa}{3E}

      \let\leq=\leqslant \let\le=\leqslant     
            
    \fi     
  \fi     
\fi 
     
\ifnfssone     
  \newmathalphabet{\mathit}     
  \addtoversion{normal}{\mathit}{cmr}{m}{it}     
  \addtoversion{bold}{\mathit}{cmr}{bx}{it}     
  \newmathalphabet{\mathbfit} 
  \addtoversion{normal}{\mathbfit}{cmr}{bx}{it}     
  \addtoversion{bold}{\mathbfit}{cmr}{bx}{it}     
  \newmathalphabet{\mathbfss} 
  \addtoversion{normal}{\mathbfss}{cmss}{bx}{n}     
  \addtoversion{bold}{\mathbfss}{cmss}{bx}{n}     
  \ifAMStwofonts     
    \ifCUPmtlplainloaded \else     
      \UseAMStwoboldmath     
      \makeatletter     
      \new@mathgroup\upmath@group     
      \define@mathgroup\mv@normal\upmath@group{eur}{m}{n}     
      \define@mathgroup\mv@bold\upmath@group{eur}{b}{n}     
      \edef\UPM{\hexnumber\upmath@group}     
      \new@mathgroup\amsa@group     
      \define@mathgroup\mv@normal\amsa@group{msa}{m}{n}     
      \define@mathgroup\mv@bold\amsa@group{msa}{m}{n}     
      \edef\AMSa{\hexnumber\amsa@group}     
      \makeatother     
      \mathchardef\upi="0\UPM19     
      \mathchardef\umu="0\UPM16     
      \mathchardef\upartial="0\UPM40     
      \mathchardef\leqslant="3\AMSa36     
      \mathchardef\geqslant="3\AMSa3E     

      \let\leq=\leqslant \let\le=\leqslant     

    \fi     
  \fi     
\fi 
     
\ifnfsstwo     
  \DeclareMathAlphabet{\mathbfit}{OT1}{cmr}{bx}{it}     
  \SetMathAlphabet\mathbfit{bold}{OT1}{cmr}{bx}{it}     
  \DeclareMathAlphabet{\mathbfss}{OT1}{cmss}{bx}{n}     
  \SetMathAlphabet\mathbfss{bold}{OT1}{cmss}{bx}{n}     
  \ifAMStwofonts     
    \ifCUPmtlplainloaded \else     
      \DeclareSymbolFont{UPM}{U}{eur}{m}{n}     
      \SetSymbolFont{UPM}{bold}{U}{eur}{b}{n}     
      \DeclareSymbolFont{AMSa}{U}{msa}{m}{n}     
      \DeclareMathSymbol{\upi}{0}{UPM}{"19}     
      \DeclareMathSymbol{\umu}{0}{UPM}{"16}     
      \DeclareMathSymbol{\upartial}{0}{UPM}{"40}     
      \DeclareMathSymbol{\leqslant}{3}{AMSa}{"36}     
      \DeclareMathSymbol{\geqslant}{3}{AMSa}{"3E}     

      \let\leq=\leqslant \let\le=\leqslant     

    \fi     
  \fi     
\fi 
     
\ifCUPmtlplainloaded \else     
  \ifAMStwofonts \else 
    \def\upi{\pi}     
    \def\umu{\mu}     
    \def\upartial{\partial}     
  \fi     
\fi

\title{A polarized synchrotron template for CMBP experiments after WMAP data}     
\author[G. Bernardi et al.]     
       {G. Bernardi$^{1,2,3}$, E.~Carretti$^1$, R.~Fabbri$^4$, C.~Sbarra$^1$,     
        S.~Poppi$^5$, S.~Cortiglioni$^1$, \and J.L.~Jonas$^6$\\     
        $^1$I.A.S.F./C.N.R. Bologna, Via Gobetti 101,      
                 I-40129 Bologna, Italy\\     
        $^2$Dipartimento di Astronomia, Universit\`a degli Studi di Bologna,      
           Via Ranzani 1, I-40127 Bologna, Italy\\     
        $^3$ATNF/CSIRO, P.O. BOX 76, EPPING, NSW, 1710, Australia\\
        $^4$Dipartimento di Fisica, Universit\`a di Firenze, Via Sansone 1,     
            I-50019 Sesto Fiorentino (FI), Italy\\     
        $^5$I.R.A./C.N.R. Bologna, Via Gobetti 101,      
                 I-40129 Bologna, Italy\\     
         $^6$Department of Physics \& Electronics, Rhodes University, PO Box 94,
        Grahamstown 6140, South Africa\\
        }     
\date{Accepted.
      Received;
      in original form}     
     
\pagerange{\pageref{firstpage}--\pageref{lastpage}}     
\pubyear{0000}     
     
\begin{document}     
     
\maketitle     
     
\label{firstpage}     
     
\begin{abstract}     
We build template maps for the polarized Galactic--synchrotron emission on large
angular scales (FWHM~=~7$^\circ$), in the 20-90~GHz microwave range, by using
WMAP data. The method, presented in a recent work, requires a synchrotron total
intensity survey and the {\it polarization horizon} to model the polarized
intensity and a starlight polarization map to model polarization angles. The
basic template is obtained directly at 23~GHz with about 94\% sky--coverage by
using the synchrotron map released by the WMAP team. Extrapolations to 32, 60
and 90~GHz are performed by computing a synchrotron spectral index map, which
strongly reduces previous uncertainties in passing from low (1.4~GHz) to
microwave frequencies. Differing from low frequency data, none of our templates
presents relevant structures out of the Galactic Plane. Our map at 90~GHz
suggests that the synchrotron emission at high Galactic latitudes is low enough
to allow a robust detection of the $E$--mode component of the cosmological
signal on large--scale, even in models with low--reionization ($\tau = 0.05$).
Detection of the weaker $B$--mode on the largest scales ($\ell < 10$) might be
jeopardized unless the value $\tau = 0.17$ found by WMAP is confirmed, and  $T/S
> 0.1$. For lower levels of the gravitational--wave background the $B$--mode
seems to be accessible only at the $\ell \sim 100$ peak and in selected
low--synchrotron emission areas.
\end{abstract}     
     
     
\section{Introduction} \label{intro}        

Recent results by  DASI (Kovac et al. 2002) and WMAP (Bennett et al 2003a) 
have opened a new window in studying both the Cosmic Microwave Background (CMB) and 
the Galaxy. 

The polarized signal detected with the DASI experiment, as well as the WMAP
measurement of the temperature--polarization cross--spectrum $C^{TE}_\ell$,
emphasized the importance of studying CMB Polarization (CMBP). In particular,
the discovery of a relevant signal on large angular scales performed by WMAP has
provided evidence of an unexpectedly early reionization era (Kogut et al. 2003).

Furthermore, WMAP provided the first all--sky total--intensity maps of the
microwave Galactic emission, allowing an insight into synchrotron, dust and
free-free components of the Galaxy (Bennett et al. 2003b).

A sound measurement of CMBP requires good knowledge of polarized foregrounds
(both Galactic and extragalactic), which are potentially more dangerous than in
total intensity. This is even more true for the $B$--mode,
retaining information on the gravitational--wave background (Kamionkowski \&
Kosowski, 1998), whose emission level is expected to be orders of magnitude below
that of the $E$-mode.

Among all foreground components,
the Galactic synchrotron radiation is expected to be 
the most relevant up to 100~GHz. The lack of data in both the CMB frequency range
and at high Galactic latitudes makes the building of templates 
necessary (Kogut \& Hinshaw 2000, Giardino et al. 2002, Bernardi et al. 2003a,
hereafter B03). These can substantially help in studying how to extract CMB maps
from contaminated signals (e.g. see Bennett et al. 2003b). 

In B03 we presented a method modelling the Galactic polarized--synchrotron
emission by using the radio--continuum total intensity surveys available at low
frequency ($< 1.4$~GHz) to model the polarized intensity, and starlight
polarization optical data as a template for polarization angles.  

Template maps obtained with this method are virtually free of Faraday rotation.
However, when extrapolating low frequency data (the only available before the
latest WMAP release) to the cosmological window, uncertainties arise on the 
spectral index to be used. In B03 we used the mean spectral index derived by Platania
et al. (1998) up to 19~GHz, since no information were available at higher
frequencies. In addition, variations across the
sky were too poorly known to be taken into account.  

The WMAP release of total--intensity synchrotron maps in the cosmological
window gives us the possibility to significatively improve our templates. 
Actually, the application of our method directly at frequencies of interest
for CMBP fully avoids extrapolation uncertainties.

Our templates are built with FWHM~$=7^{\circ}$ at the three frequencies of 23,
32 and 90~GHz interesting for the SPOrt experiment (Cortiglioni et al. 2004),
one of whose aims is studying the CMBP foregrounds. We also generate a template
at 60~GHz, this frequency being of interest for other CMB experiments (WMAP,
Planck). All of the templates cover large fractions of the sky, namely 94\% at
23~GHz and 92\% at 32, 60 and 90~GHz. In addition, we calculate
polarized--syncrotron angular power spectra both for the full-sky templates and
for the high Galactic latitude emission, discussing possible effects on the
measurement of CMBP $E$ and $B$--mode spectra.

The paper is organized as follows: in Section~\ref{model} we apply our method to
WMAP  data at 23~GHz and provide a Galactic polarized synchrotron template at
the same frequency; in Section~\ref{extr} we extrapolate our results up to
90~GHz; in Section~\ref{tot_aps} we describe the angular power spectra of our
full--sky templates at the different frequencies, in Section~\ref{for_aps} we
compute the angular power spectra for the high Galactic latitude emission and
compare synchrotron to CMBP and, finally, conclusions are presented in
Section~\ref{conc}.

\section{A polarized template at 23~GHz using WMAP data} 
\label{model}     
            
In B03 we presented a method to obtain a template for the polarized 
Galactic synchrotron emission based on total intensity radio maps, starlight
polarization data and the modellization of a {\it polarization horizon} as
suggested by several authors (Duncan et al. 1997, Gaensler et al. 2001,
Landecker et al. 2002).

The method is based on separate derivations of the polarized intensity and
polarization angle maps. Whereas the latter is obtained from starlight
polarization data, the former is the result of the application to
total--intensity synchrotron maps of a filter mimicking the effects of the
polarization horizon. 

The main steps of our method can be summarized as follows:
 \begin{enumerate}
 \item{} a total intensity synchrotron map is obtained from
        available surveys, after separating the CMB and free-free components
        when necessary;
 \item{} a filter accounting for the effects of the polarization horizon is
 	applied to the $I$ map, providing the polarized intensity $I_p =
	\sqrt{Q^2 + U^2}$;
 \item{} a polarization--angle map is obtained from starlight data by
         interpolating the non-uniform data coverage;
 \item{} the polarized--intensity and  polarization--angle maps are used to
 	compute  $Q$ and $U$ templates including a convolution with a
         FWHM~$=7^\circ$ Gaussian filter to match the SPOrt angular resolution
         (finer resolutions are not possible due to the sampling
          of starlight polarization data). 
 \end{enumerate} 

Before WMAP, total intensity surveys were available up to 1.4~GHz and our templates
were limited by our ignorance of the spectral index distribution needed to
extrapolate to higher frequencies. Indeed, previous templates at 22, 32, 60 and
90~GHz were obtained by using a constant spectral index $\beta = -3$,
corresponding to the mean value between 1.4 and 19~GHz (Platania et al. 1998).

The model is now substantially improved by using the latest WMAP data, including 
a total--intensity synchrotron map (cleaned from
CMB, free--free and dust contributions) with resolution of $1^\circ$  at 23~GHz,
that directly allows us to compute the template at a frequency interesting 
for CMBP purposes.

The polarized intensity map is obtained by applying to the 23~GHz synchrotron map
the equation
\begin{equation}
I_p(\nu,l,b) = p \frac{R_{ph}}{L(l,b)}I(\nu,l,b)
\end{equation}
mimicking the polarization horizon effects (see B03). Here 
$I_p$ and $I$ are the polarized and total intensity of the synchrotron
radiation, respectively, $L(l,b)$ is the thickness of the synchrotron emitting
region, $p$ is the polarization degree, $R_{ph}$ is the distance of the 
polarization horizon. We adopt the same normalization factor $p \, R_{ph}
= 0.9$~Kpc as computed in B03.

\begin{figure*}      
 \begin{center}  
 \includegraphics[width=0.4\hsize,angle=90]{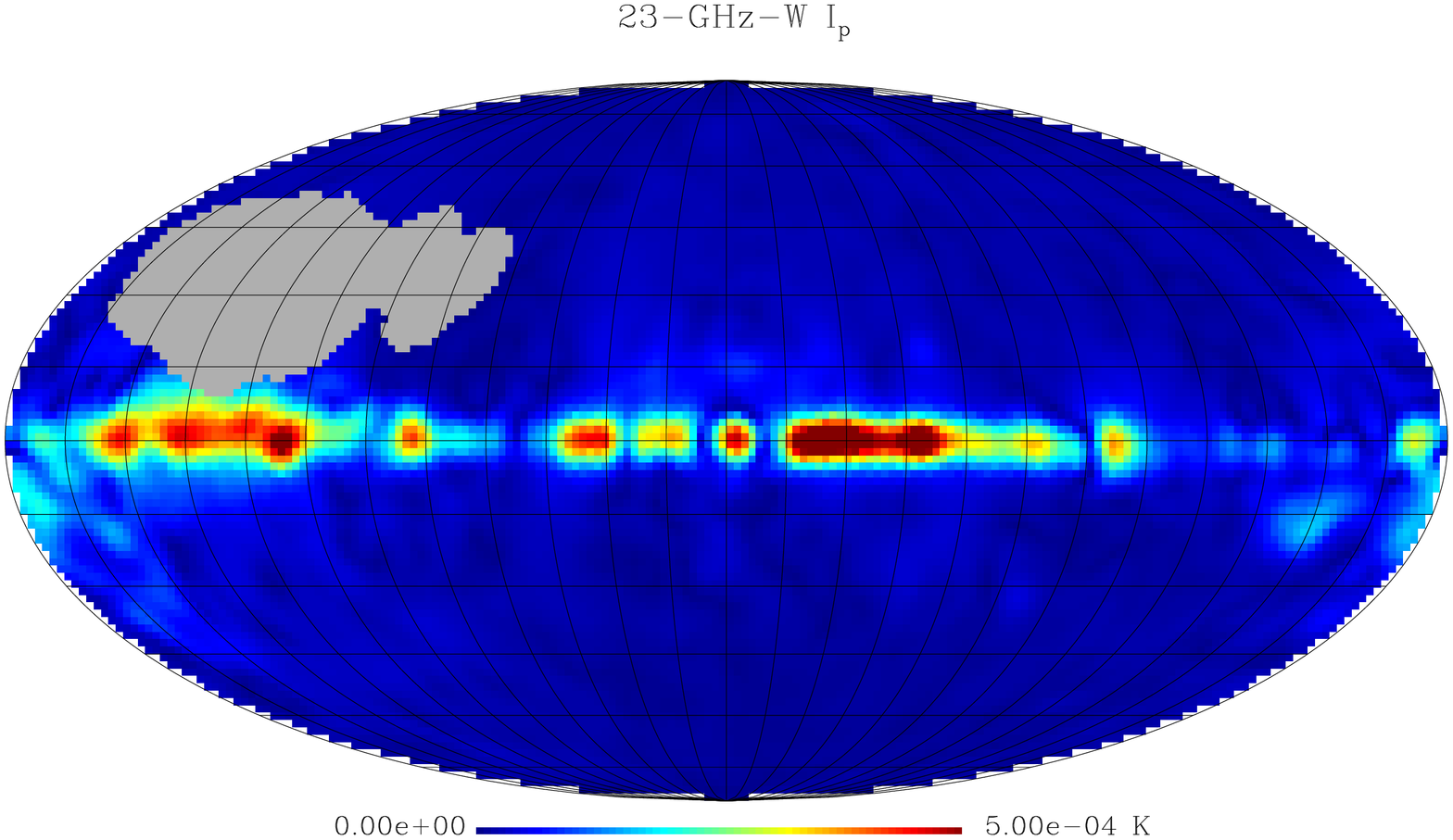}      
 \includegraphics[width=0.4\hsize,angle=90]{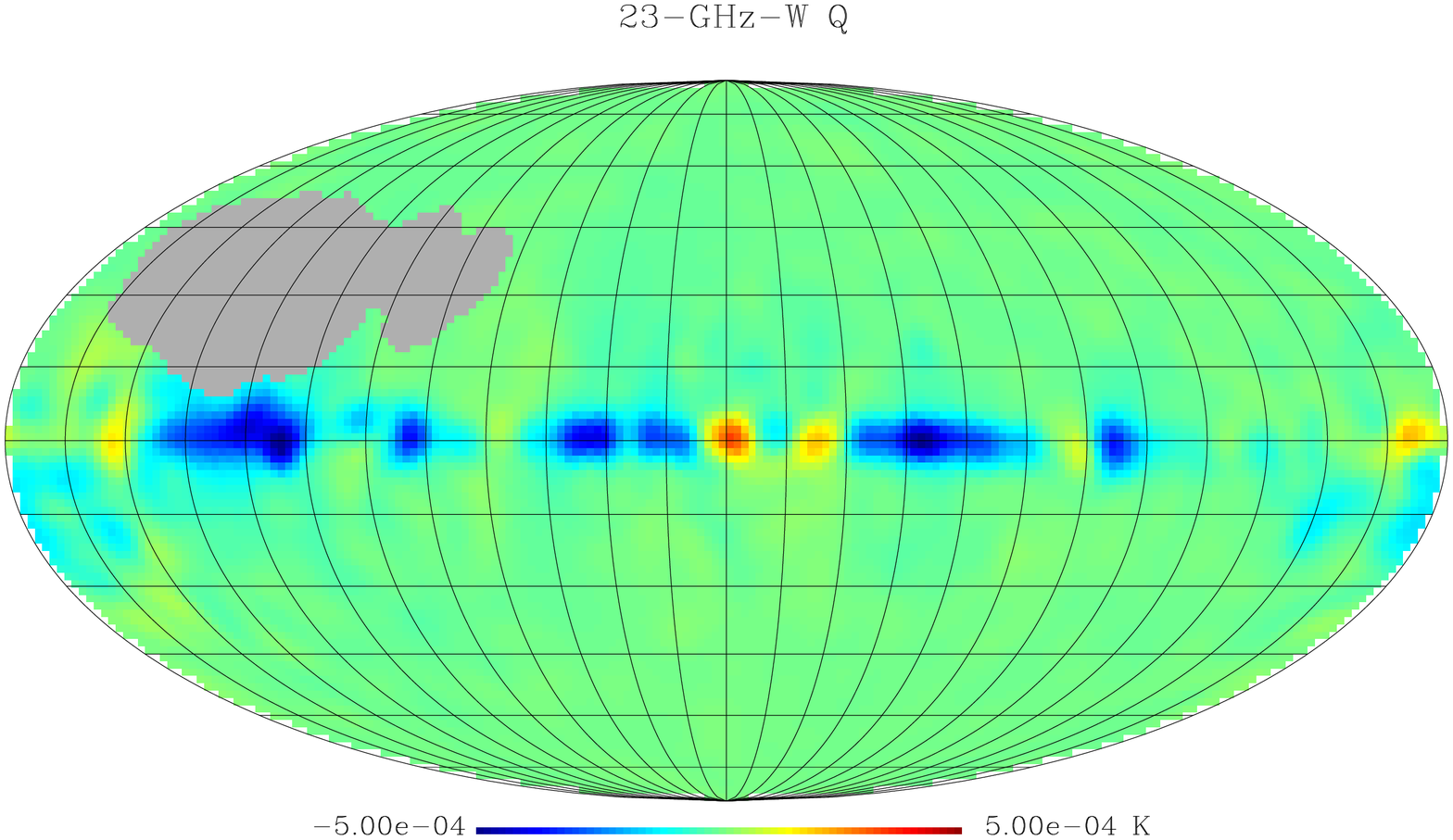}      
 \includegraphics[width=0.4\hsize,angle=90]{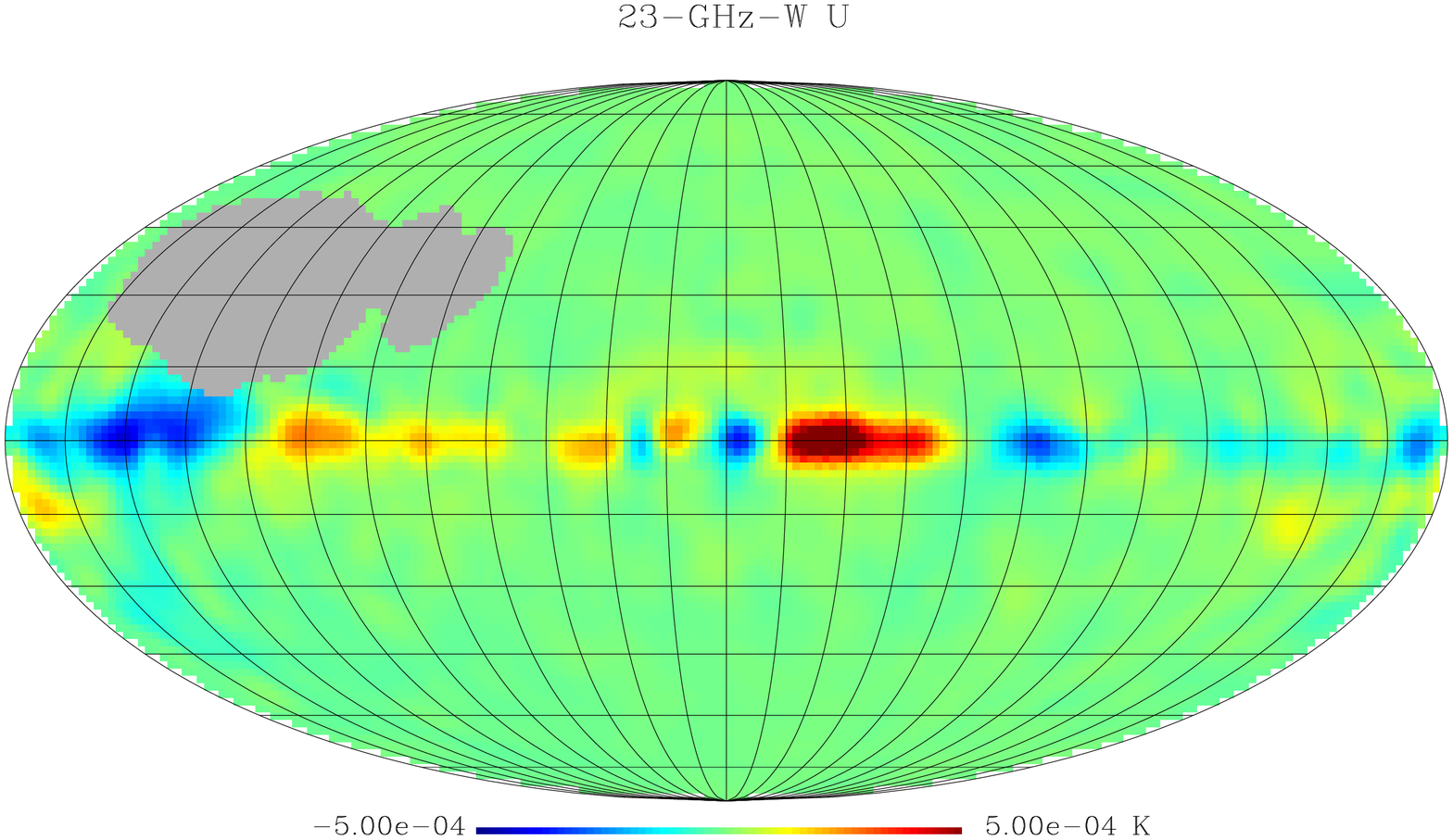}
 \caption{Our 23~GHz polarized synchrotron template obtained using WMAP data as
 total intensity emission. Maps are convolved with a FWHM = $7^\circ$ Gaussian
 filter.}
 \label{wmap_23}     
 \end{center}  
 \end{figure*}

From this polarized brightness temperature, and using the same 
polarization--angle map as in B03, $Q$, $U$ and $I_p$ maps are obtained. 
They are shown in Figure
\ref{wmap_23}, where the grey region corresponding to the North Celestial Pole
 is an area where the interpolation of starlight polarization angles failed
(see B03 for details). It is worth emphasizing that the new maps cover
almost 94\% of the sky, to be compared with the half--sky coverage of
the B03 template.

\begin{figure}      
 \begin{center}  
 \includegraphics[width=0.6\hsize,angle=90]{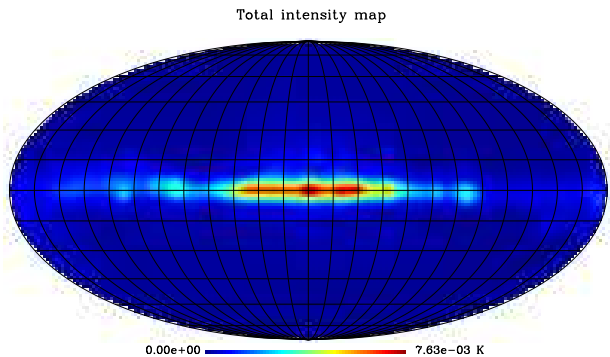}      
 \caption{WMAP's total intensity synchrotron emission at 23~GHz (from Bennett et
 al. 2003b) smoothed with a $7^\circ$ beam.}
 \label{tot_int}     
 \end{center}  
 \end{figure}

\begin{figure}      
 \begin{center}  
 \includegraphics[width=1.0\hsize]{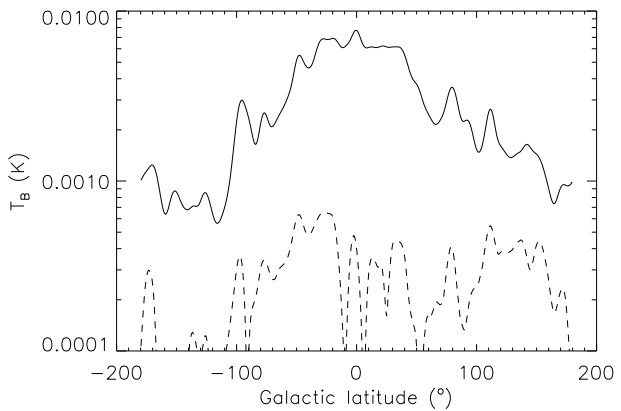}
 \caption{Comparison between the total (observed, solid line) and polarized
 (template, dashed line) brightness temperatures along the Galactic Plane at
 23~GHz. We consider all the pixels with $|b| \le 0.1^\circ$} 
 \label{profile_fig}     
 \end{center}  
 \end{figure}

Confirming the B03 results, the distributon of the polarized intensity is quite
different from that of total intensity (Figure \ref{tot_int}). The latter
peaks towards the Galactic centre and traces well the Galactic Plane. Instead,
the polarized emission has a sub-dominant Galactic Centre and the Galactic Plane
shows a patchy structure (Figure \ref{profile_fig}). 

\begin{figure}      
 \begin{center}  
 \includegraphics[width=0.6\hsize,angle=90]{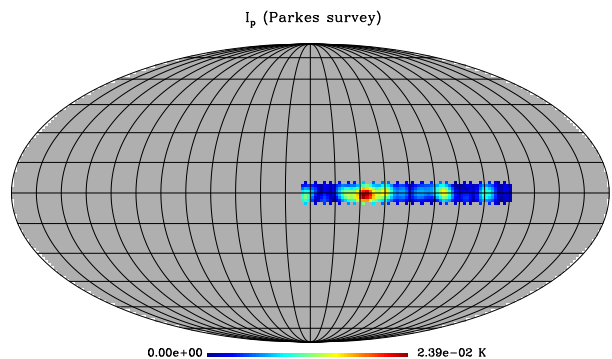}      
 \caption{Polarized intensity emission from the Parkes survey at 2.4~GHz smoothed on
 $7^\circ$ (Duncan et al. 1997).}     
 \label{parkes_pi}     
 \end{center}  
 \end{figure}

A comparison with the B03 template shows a general agreement between the two
maps along the Galactic Plane, and in particular in the following three regions: 
\begin{enumerate}    
 \item{} {\it Fan region} located at $100^\circ \leq l \leq 160^\circ$;
 \item{} the region at $30^\circ \leq l \leq 45^\circ$;
 \item{} the region around $ l \simeq 15^\circ$, though only partially present
 in the B03 template.
\end{enumerate}     
A comparison with the Parkes survey (Figure \ref{parkes_pi}) shows remarkable
similarities as well. The brightest polarized region in our template is at
Galactic longitude $315^\circ \le l \le 345^\circ$, where also Duncan et al.
(1997) found a high level in the polarized background at 2.4~GHz (see also
Figure \ref{parkes_pi}). The bright features present in our template correspond
to emission spots in the Duncan map, with the exception of the large region
centred at $l \sim 300^\circ$. This discrepancy is the same noted in B03 and is
not surprising. Actually, this is a peculiar area with strong variations in
Rotation Measures (Sofue \& Fujimoto 1983, Duncan et al. 1997, B03), so that the
absence of polarized emission in the Parkes data is likely to be due to Faraday 
depolarization.

Out of the Galactic Plane, the model computed with the WMAP data (hereafter the
23--GHz--W template) does not show any relevant structure. This differs from the
emission at 1.4~GHz, where the Northern Galactic Spur is among the main
polarized structures, both in the Brouw \& Spoelstra (1976) data and our 1.4~GHz
B03 model. The situation is similar in the total--intensity case, where the
Northern Galactic Spur is a prominent feature at 1.4~GHz, but is no more
significant at 23~GHz. This behaviour could be explained following Bennett et
al. (2003b), who showed that the spectral index of the synchrotron radiation in
the 0.408--23~GHz range is quite different in the Plane and out of it. The
Galactic Plane having a flatter spectral index becomes more and more dominant
with increasing frequency.

\section{Spectral index behaviour and extrapolation to higher frequencies} 
\label{extr}     

The most important frequency range for CMBP studies is around 90~GHz, where the
polarized foregrounds are expected to have minimum emission. Since the 94~GHz
synchrotron map provided by WMAP appears to be dominated by noise at high
Galactic latitudes, we do not apply our method directly to it to provide a
synchrotron template at 90~GHz. Instead, we prefer to extrapolate our 23~GHz
template, characterised by a better $S/N$ ratio.


Faraday rotation effects are negligible at frequencies higher than 20~GHz (e.g.
see B03). In absence of RM modulation,  polarization and total intensity are
expected to have the same behaviour.

Synchrotron spectral--index  variations across the sky have to be carefully
modelled when extrapolating syncrotron maps to higher frequencies. Bennett et
al. (2003b) provided a spectral--index map, for the 0.408--23~GHz range, which
applies to a mixture of synchrotron and free--free emissions and is not directly
useful for our purposes.

We thus extract a nearly full--sky synchrotron map at 1.4~GHz, by means of
a modified Dodelson technique (B03) to separate between components, from the
Haslam et al. (1982) survey at 0.408~GHz, the Reich (1982) survey at 1.4~GHz,
and the  Jonas et al.  (Jonas, Baart \& Nicolson, 1998) survey at 2.3~GHz.

Properties of data at 0.408~GHz and 1.4~GHz  have already been described
in B03 (and references therein). Data at 2.3~GHz correspond to a radio--continuum survey of
the Southern sky ($-83^\circ < \delta < 13^\circ$), with angular resolution of
20~arcmin.

All maps are smoothed to the same resolution of 51~arcmin, corresponding to
data at 0.408~GHz. The free--free subtraction is performed adopting the B03 model for the
spatial distribution of the synchrotron spectral index  which, being symmetric with
respect to both the Galactic Plane and the Galactic centre, can be extended to
the Southern sky.

\begin{figure}      
 \begin{center}  
 \includegraphics[width=0.6\hsize,angle=90]{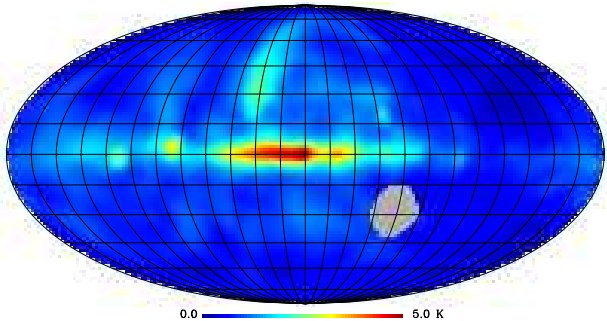}      
 \caption{Total intensity synchrotron map at 1.4~GHz on $7^\circ$ (cleaned from
 free--free contribution) obtained from Haslam et al. (1982), Reich (1982) and
 Jonas et al. (1998) maps.}
 \label{tot_int_1400MHz}     
 \end{center}  
 \end{figure}

Due to different sky coverages of the available surveys, synchrotron maps are
built separately in the Northern and in the Southern emispheres. Where the two
maps overlap, we retain information from the Northern one, since the zero level
of its parent surveys are more accurately determined than that at 2.3~GHz.
Moreover, we correct the offset of the Southern synchrotron map shifting its
mean level to the that of the Northern map in the overlapping region.
Discontinuities at borders are at negligible levels for our purposes, being
cancelled out by  the convolution with a FWHM~$=7^\circ$ Gaussian filter. The
final map, covering almost 99\% of the sky, is shown in Figure
\ref{tot_int_1400MHz}.

\begin{figure}      
 \begin{center}  
 \includegraphics[width=0.6\hsize,angle=90]{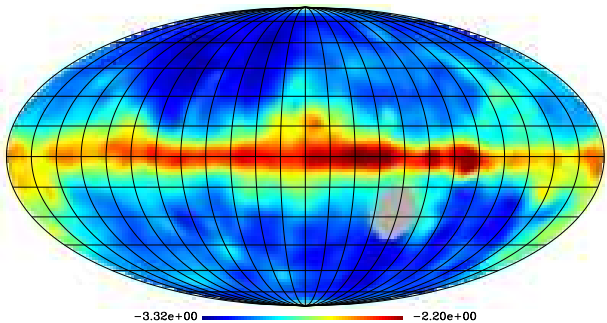}      
 \caption{Spectral index map of the synchrotron emission on $7^\circ$ angular
 scales between 1.4 and 23~GHz. The grey region around the South Celestial Pole
 is the unobserved area in the 2.3~GHz survey.}     
 \label{spectral_index}     
 \end{center}  
 \end{figure}

A comparison with the 23~GHz synchrotron map provided by WMAP allows us to
determine the synchrotron spectral index in the 1.4--23~GHz range
(Figure~\ref{spectral_index}).
 \begin{figure}      
 \begin{center}  
 \includegraphics[width=0.6\hsize,angle=90]{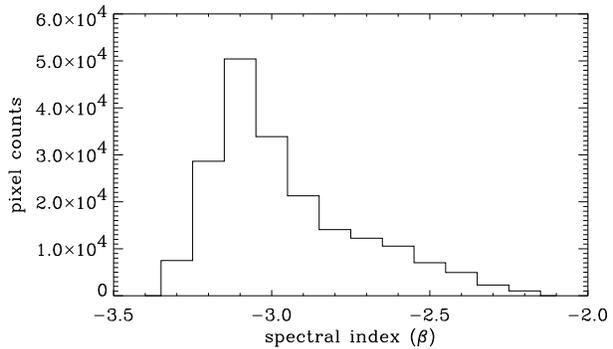}      
 \caption{Synchrotron spectral index distribution between 1.4 and 23~GHz.}      
 \label{spectral_histogram}     
 \end{center}  
 \end{figure}                      

The overall pattern of our synchrotron spectral--index map is substantially in
agreement with that by  Bennett et al. (2003b): the spectral index is rather
flat along the Galactic Plane, whereas steeper spectra are found at high
Galactic latitudes. As shown in Fig.\ref{spectral_histogram}, the typical
spectral index is $\beta =-3.1$, which is steeper than the value found by
Bennett et al. (2003b); this might be due to the thermal contribution in their
data.

\begin{figure*}      
 \begin{center}  
 \includegraphics[width=0.4\hsize,angle=90]{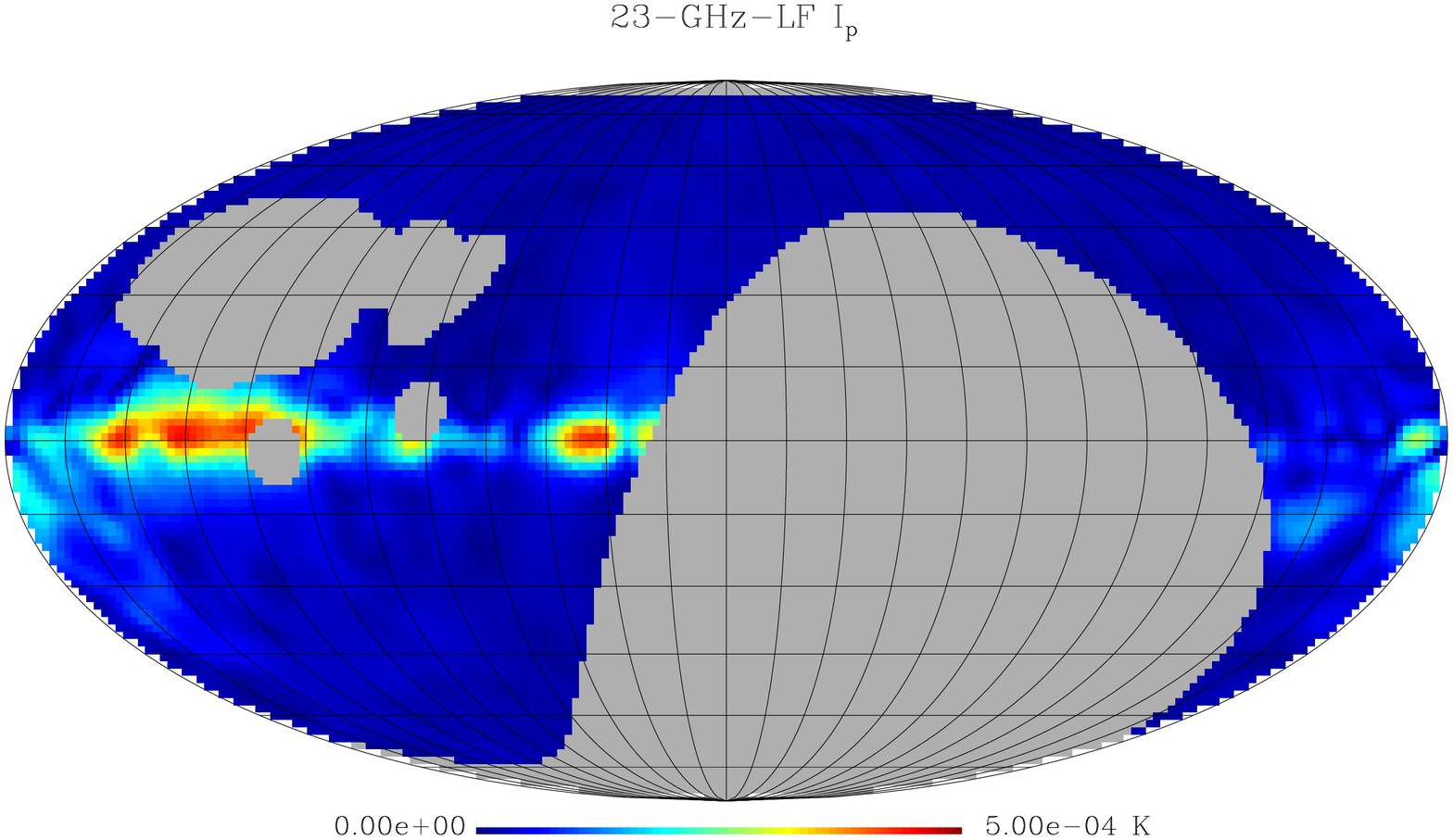}      
 \includegraphics[width=0.4\hsize,angle=90]{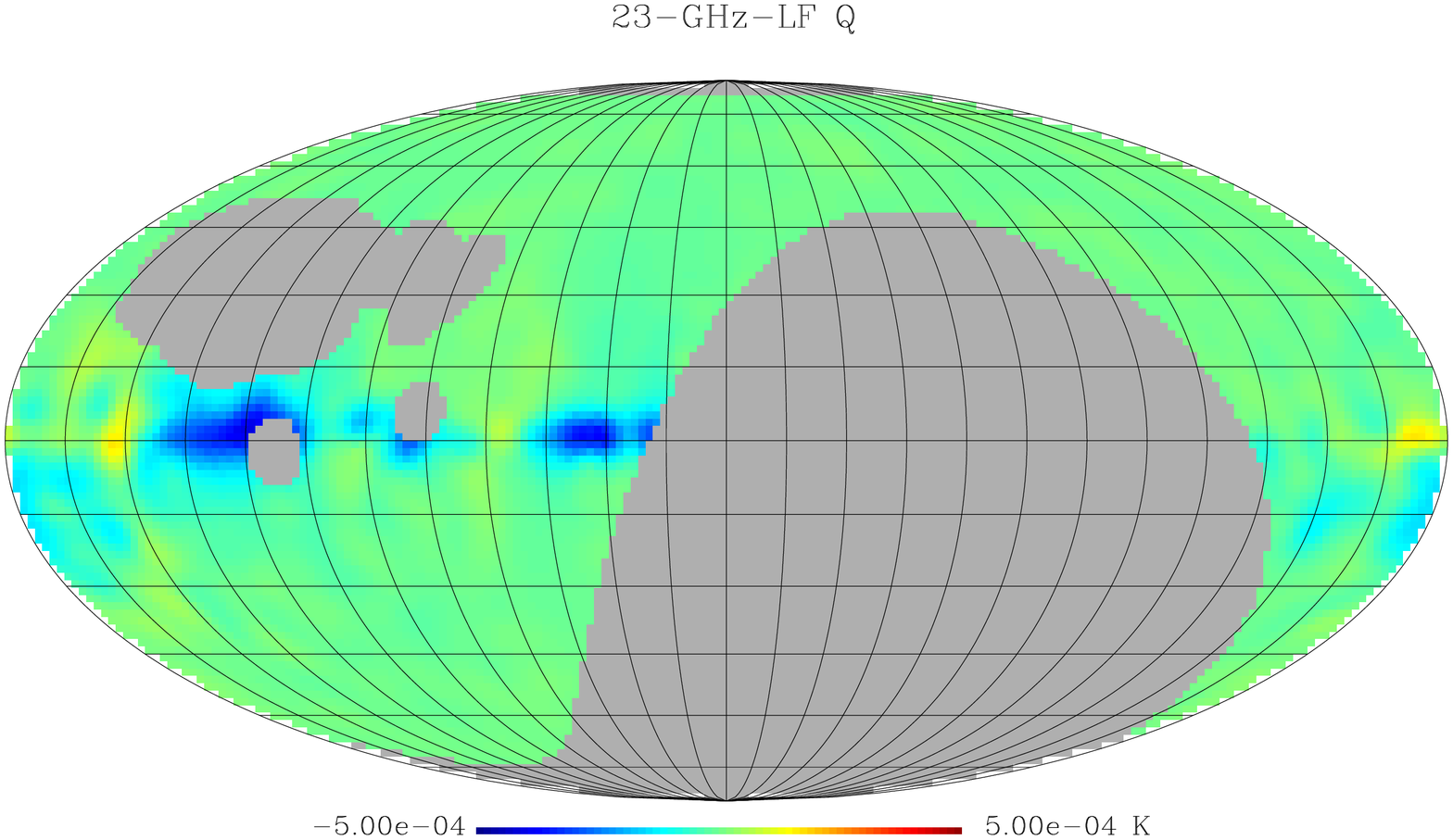}      
 \includegraphics[width=0.4\hsize,angle=90]{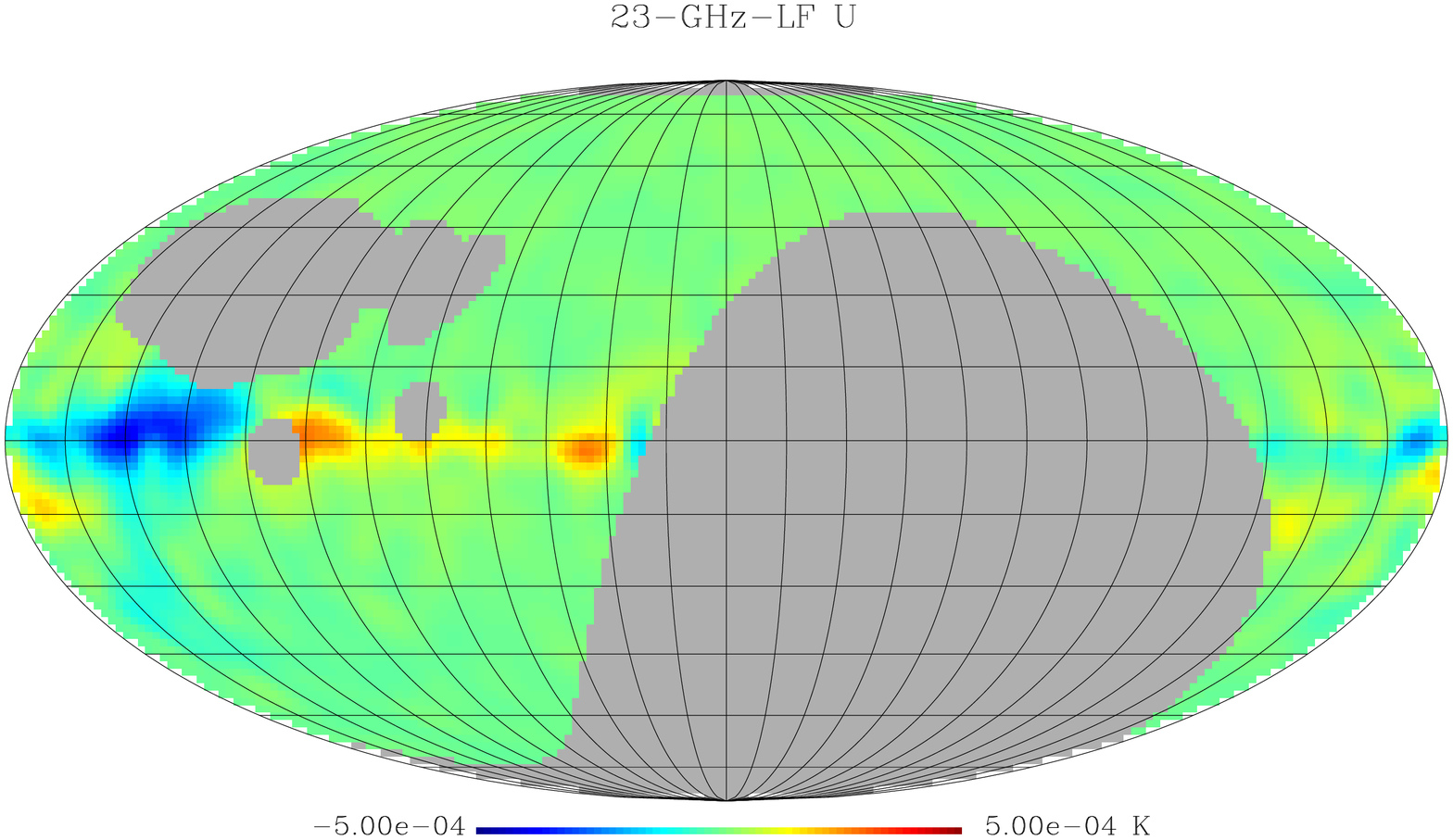}
 \caption{Maps of our template obtained from the 1.4~GHz B03 one scaled to
 23~GHz (23--GHz--LF). The angular resolution is $7^\circ$.}      
 \label{our_model}     
 \end{center}  
 \end{figure*}

We use our spectral--index map to extrapolate the 1.4~GHz B03 template 
up to 23~GHz (hereafter the 23--GHz--LF [low frequency] template). 
The result is shown in Figure~\ref{our_model}, and can be compared with
the 23--GHz--W map to further test our template--building procedure.

A general agreement in the morphology  can be observed. In particular, the most relevant 
characteristic in both maps is the lack of features at high Galactic latitudes.
The Northern Galactic Spur, present at 1.4~GHz, disappears when extrapolated 
with the spectral--index map of Figure \ref{spectral_index}, in 
agreement with 23--GHz--W. Also the main features in the Galactic Plane are 
similar.

Indeed, the model applied to the 0.408--1.4~GHz data (including the free--free
component separation) and extrapolated to 23~GHz by using spatially varying
spectral indeces presents results very similar to the 23--GHz--W template, which
is free from extrapolation uncertainties. 
 \begin{figure*}      
 \begin{center}  
 \includegraphics[width=0.4\hsize,angle=90]{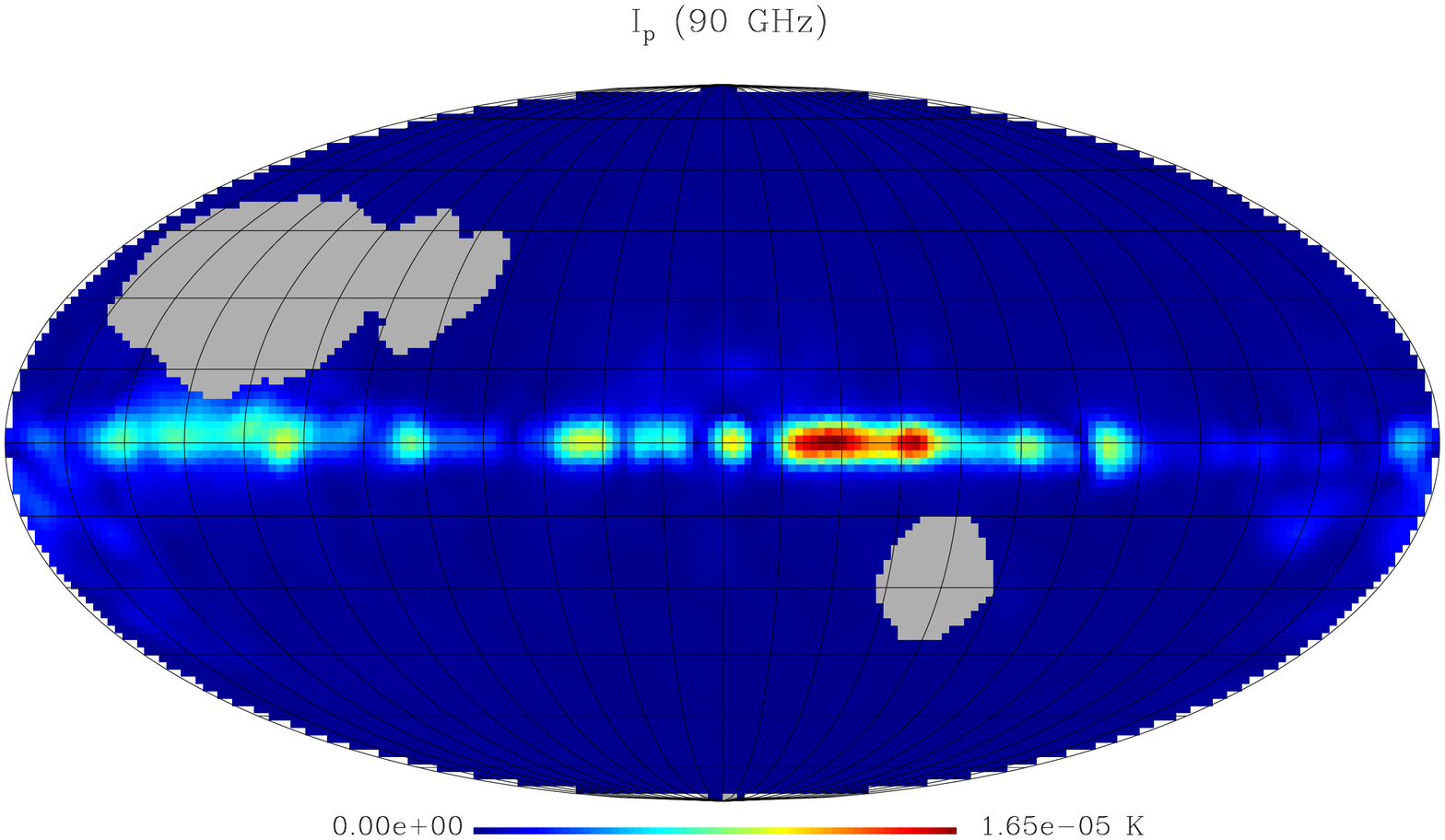}      
 \includegraphics[width=0.4\hsize,angle=90]{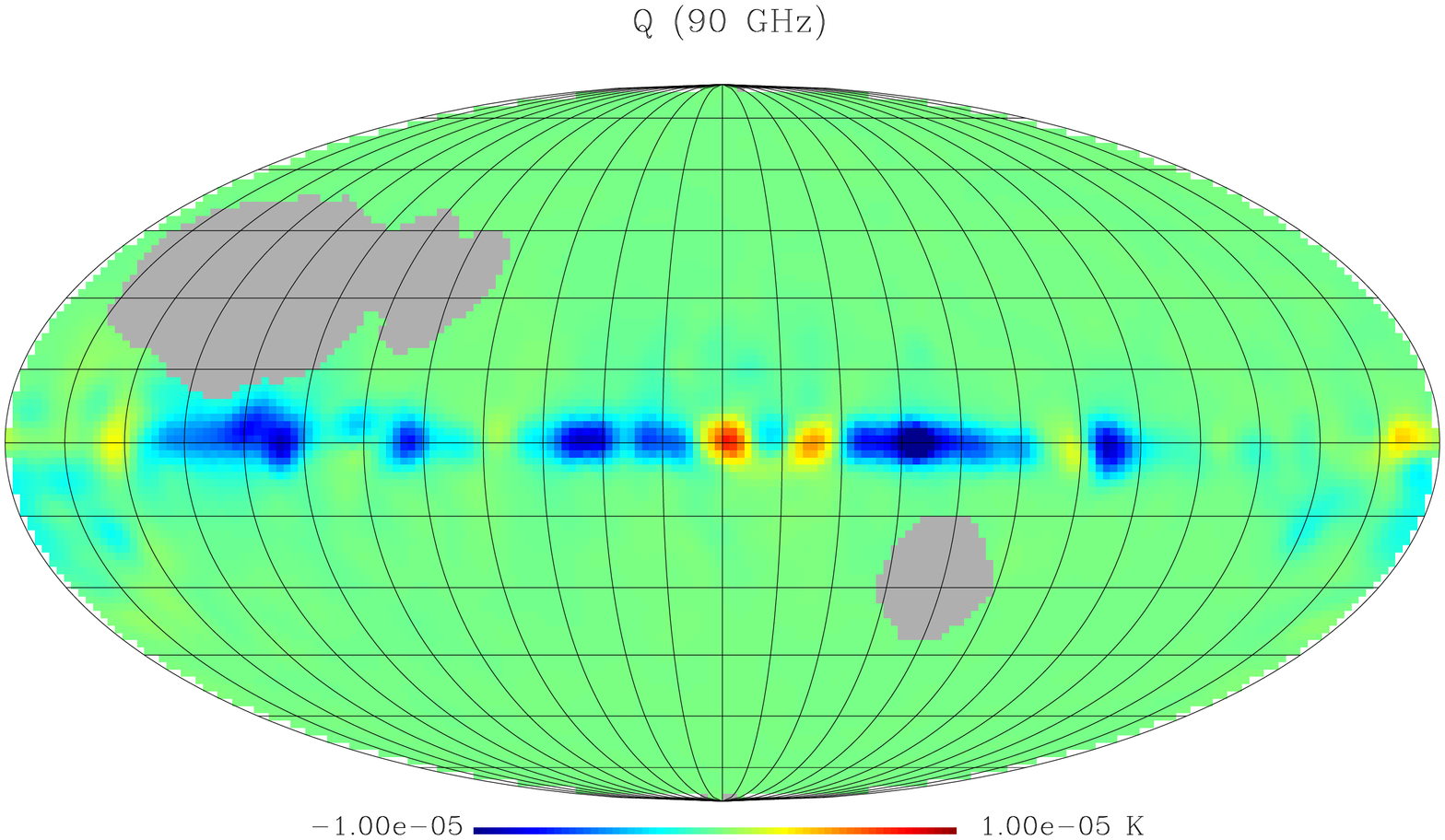}      
 \includegraphics[width=0.4\hsize,angle=90]{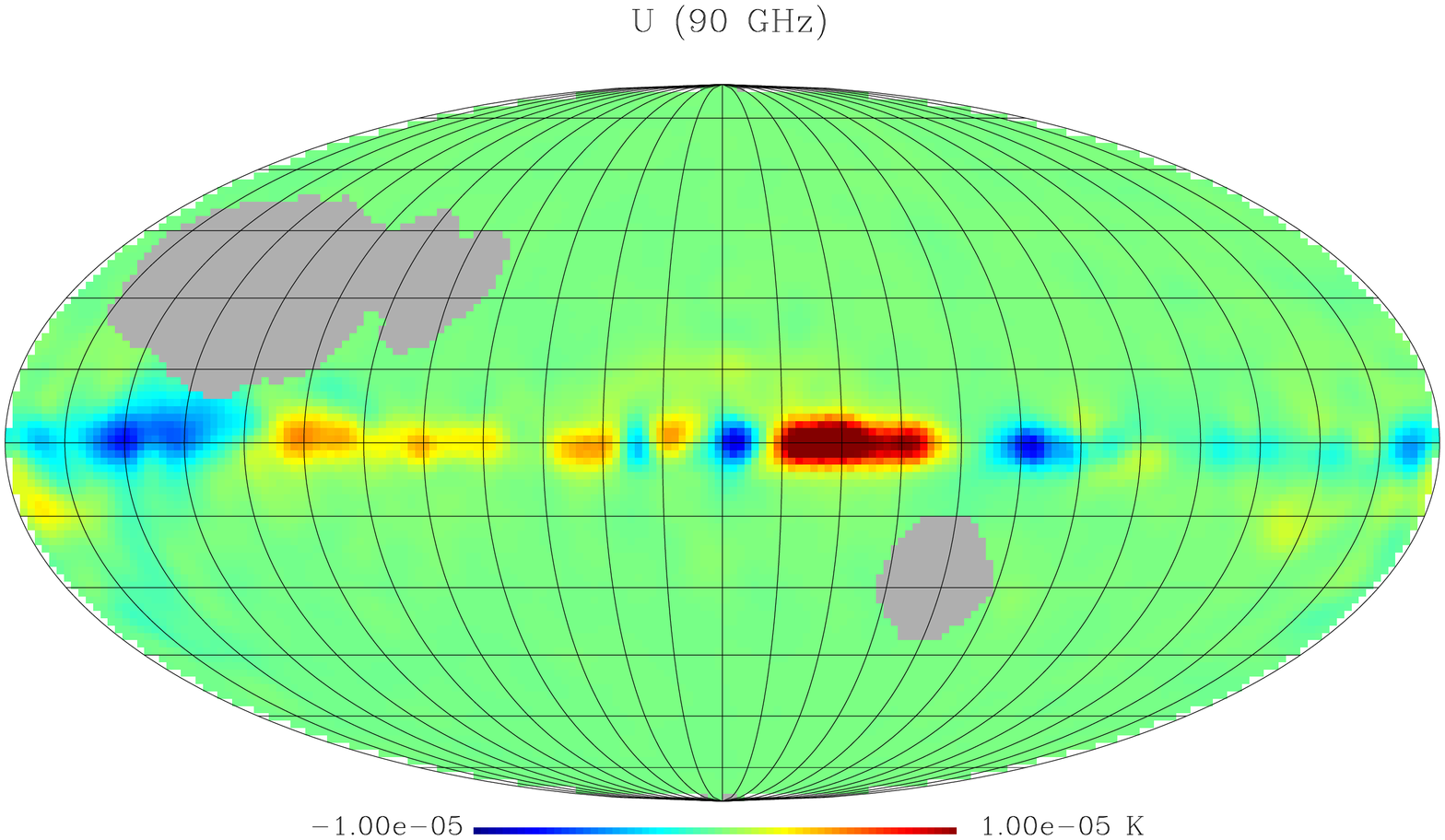}
 \caption{Maps of our template at 90~GHz. The angular resolution is $7^\circ$.}
 \label{model_90}     
 \end{center}  
 \end{figure*}                      
 \begin{figure*}      
 \begin{center}  
 \includegraphics[width=0.45\hsize]{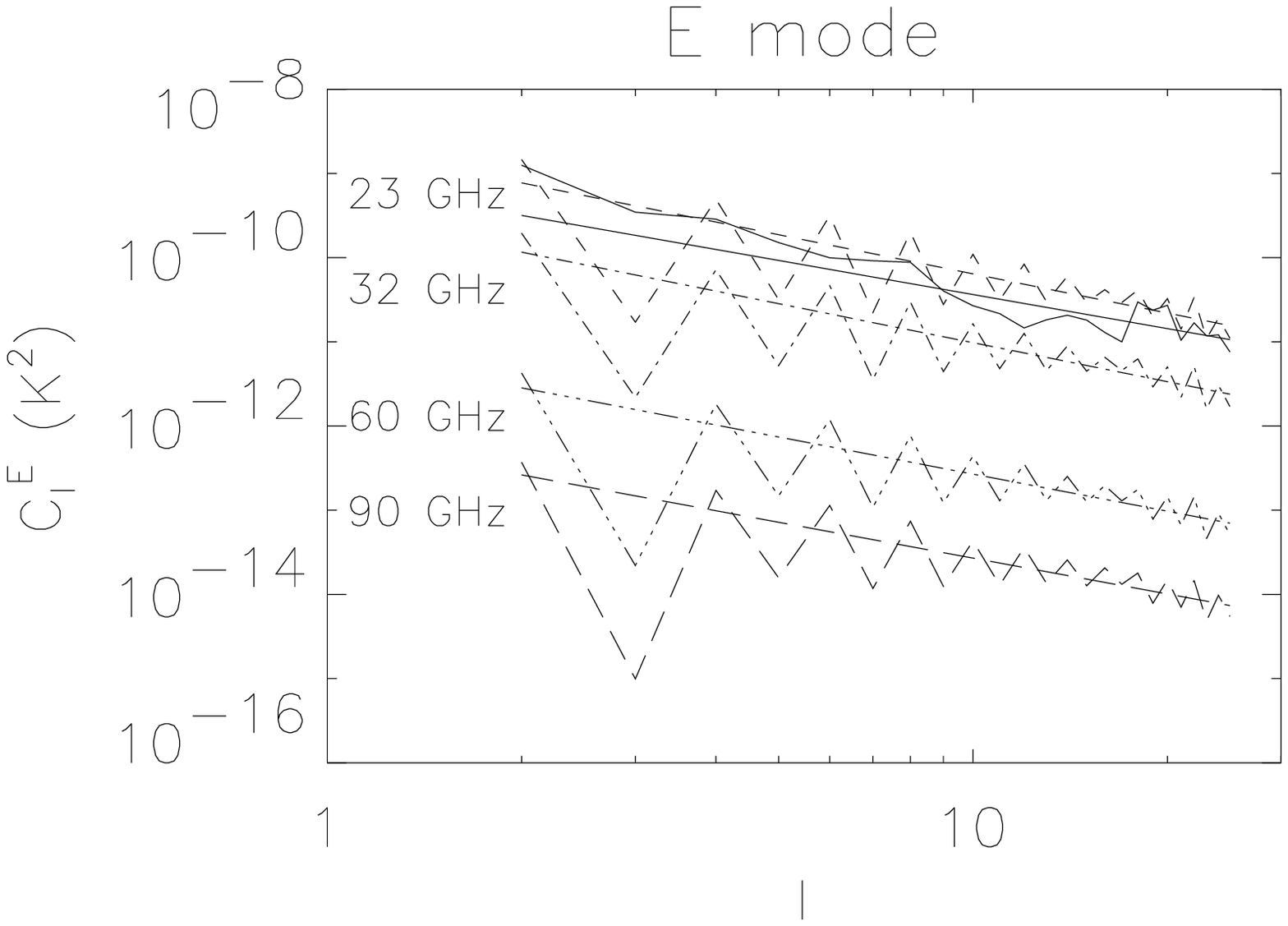}      
 \includegraphics[width=0.45\hsize]{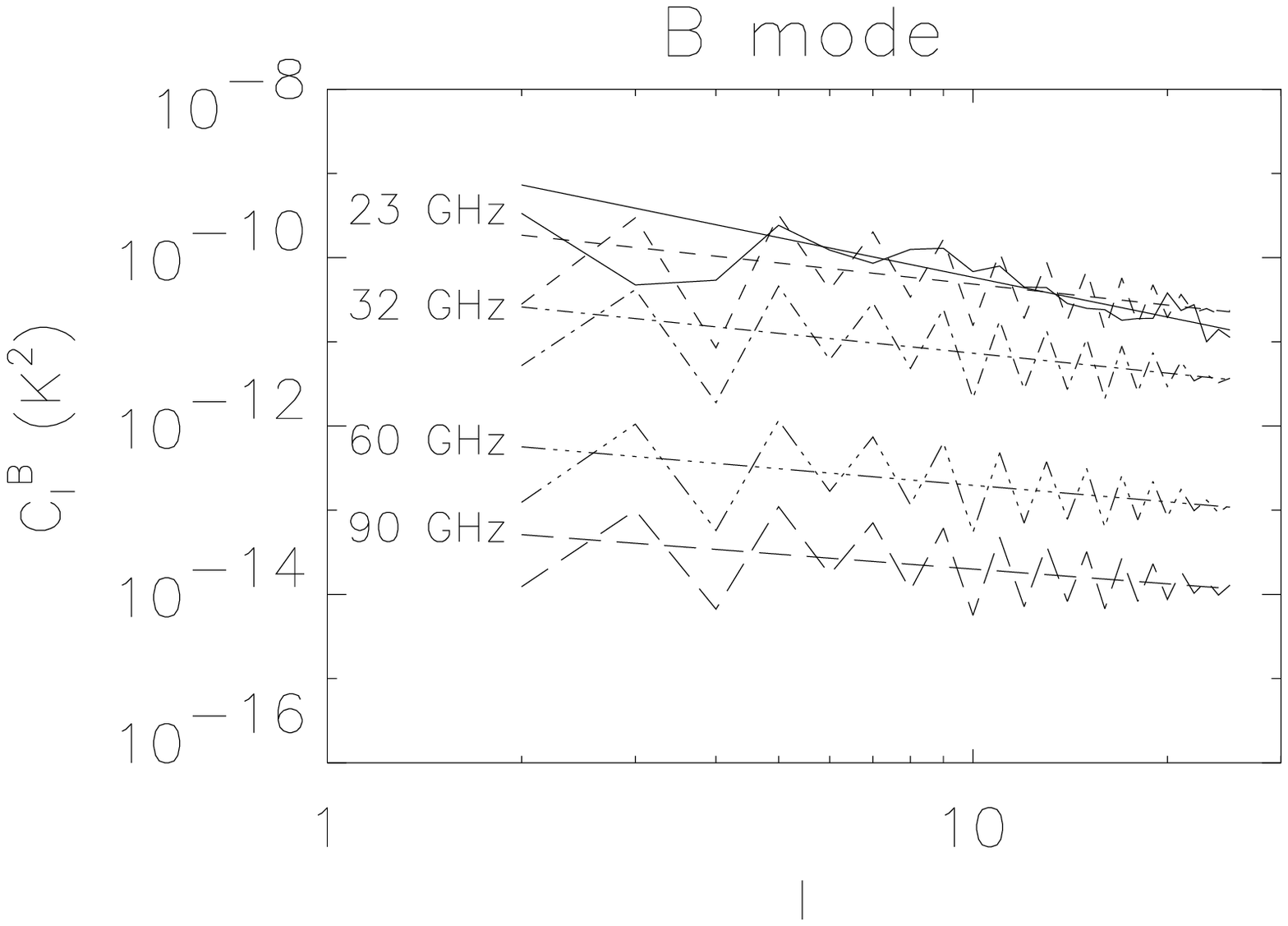}      
 \includegraphics[width=0.45\hsize]{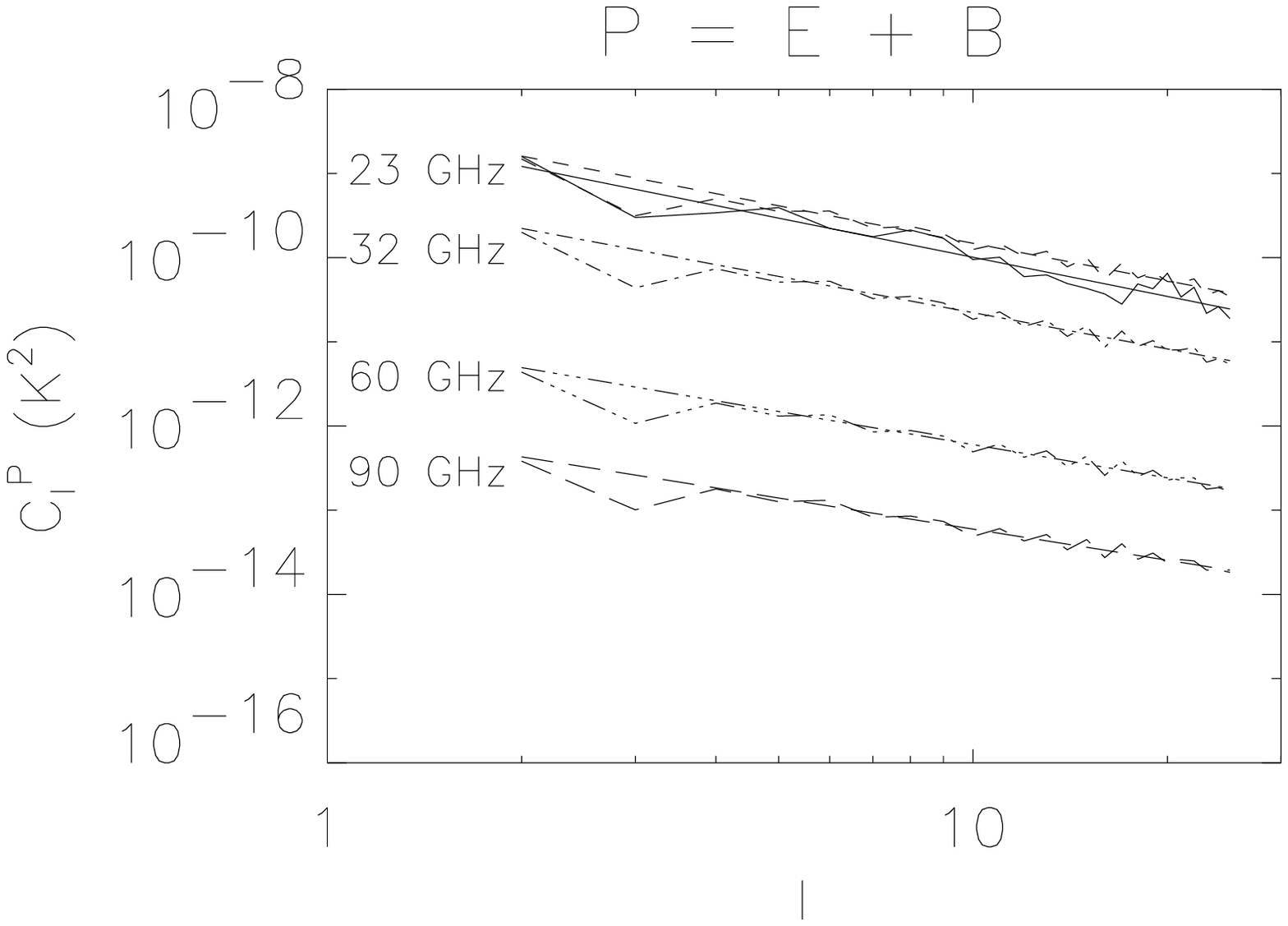} 
 \includegraphics[width=0.45\hsize]{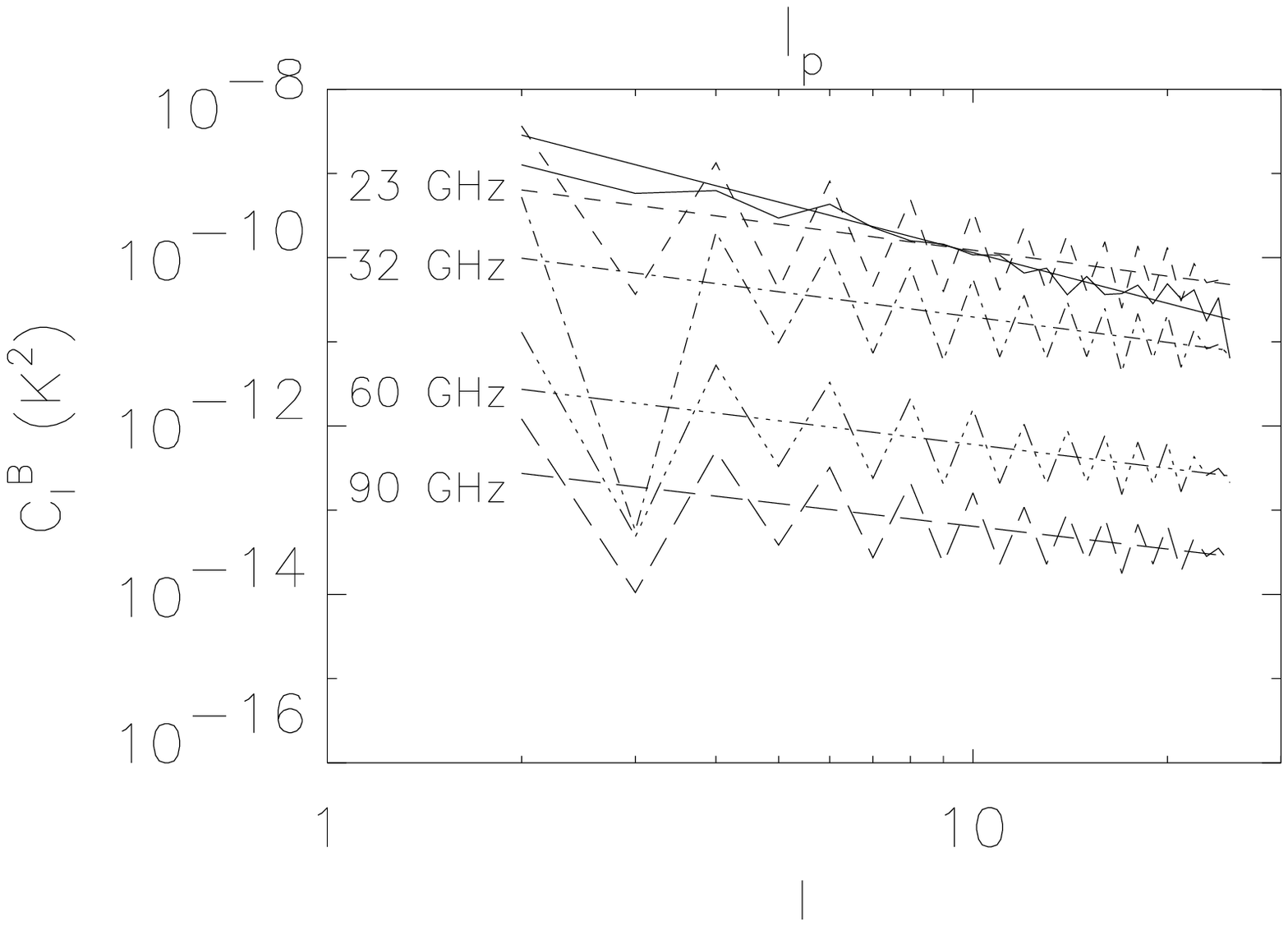}
 \caption{$C^E_\ell$, $C^B_\ell$, $C^P_\ell$ and $C^{I_p}_\ell$ spectra 
 computed for the 23--GHz--LF (solid) and 23--GHz--W templates (dotted). 
 Spectra for the 
 32~(dashed), 60~(dashed--dotted) and
 90~GHz~(dashed--triple dotted) models 
 are shown as well.}     
 \label{eb_aps}     
 \end{center}  
 \end{figure*}                      
 \begin{table*}    
 \begin{center}    
 \centering
 \caption{Best fit APS slopes and amplitudes as obtained from our templates in
 the $2 \le \ell \le 25$ range using WMAP data. First  and second lines report
 the 23--GHz--LF and 23--GHz--W templates respectively.} 
 \begin{tabular}{|p{1cm}|*{8}{c|}|}    
 $\nu$ & $\alpha_E$ & $C_{10}^E$ & $\alpha_B$ & $C_{10}^B$ &
 $\alpha_P$ & $C_{10}^P$ & $\alpha_{I_p}$ & $C_{10}^{I_p}$ \\
 (GHz) &  & $(\mu$K$^2)$ &  & $(\mu$K$^2)$ &  & $(\mu$K$^2)$ & & $(\mu$K$^2)$\\
\hline    
23--LF    & $1.34 \pm 0.09$ & $37 \pm 2$        & $1.57 \pm 0.09$ & $58 \pm 4$ 
          & $1.54 \pm 0.09$ & $101 \pm 6$       & $2.00 \pm 0.09$ & $115 \pm 7$\\     
23--W     & $1.55 \pm 0.09$ & $64 \pm 4$        & $0.83 \pm 0.09$ & $49 \pm 3$ 
          & $1.48 \pm 0.09$ & $148 \pm 9$       & $1.00 \pm 0.09$ & $123 \pm 8$\\     
32        & $1.54 \pm 0.09$ & $9.7 \pm 0.6$     & $0.79 \pm 0.09$ & $7.3 \pm 0.4$ 
          & $1.43 \pm 0.09$ & $22 \pm 1$        & $1.00 \pm 0.09$ & $20 \pm 1$\\     
60        & $1.47 \pm 0.09$ & $0.27 \pm 0.02$   & $0.65 \pm 0.09$ & $0.20 \pm 0.01$ 
          & $1.31 \pm 0.09$ & $0.60 \pm 0.04$   & $0.94 \pm 0.09$ & $0.60 \pm 0.04$\\
90        & $1.41 \pm 0.09$ & $(27 \pm 2) \times 10^{-3}$ & $0.58 \pm 0.09$ & $(20 \pm 1) \times 10^{-3}$
          & $1.23 \pm 0.09$ & $(60 \pm 4) \times 10^{-3}$ & $0.90 \pm 0.09$ & $(65 \pm 4) \times 10^{-3}$
\label{power_spectra}
\end{tabular}
\end{center}
\end{table*}

In order to produce template maps up to 90~GHz, the spectral index map in Figure
\ref{spectral_index} is not sufficient. A steepening at higher frequencies,
in fact, has been found by Bennett et al. (2003b), who reported a mean steepening
of 0.5 in the 23--41~GHz range. To extrapolate our 23--GHz--W template,
thus, we increase the spectral indexes by this amount. 

Figure~\ref{model_90} shows $I_p$, $Q$ and $U$ maps at 90~GHz, which is
the most interesting frequency for CMB measurements. Maps at 32 and 60 GHz look
qualitatively rather similar. As already pointed out in this section, the
flatter Galactic Plane component becomes more and more dominant with increasing
frequency. A further important characteristic is that the Galactic Plane
essentially retains its morphology. However, again due to the different spectral
behaviour, the {\it Fan region} is fainter and fainter with increasing
frequency respect with the $300^\circ < l < 330^\circ$ region.

\section{Angular properties of the polarized template}
\label{tot_aps}

The distribution of the polarized emission at different angular scales is 
described by the angular power spectra (APS) $C^E_\ell$, $C^B_\ell$ and
$C^P_\ell = C^E_\ell + C^B_\ell$. We compute them by the correlation--function
method (see Sbarra et al. 2003 for details), accounting for irregularities in
the sky coverage. The APS of our templates are reported in Figure~\ref{eb_aps}.
Their overall behaviour, with particular reference to $C^P_\ell$, can be
represented by a power law:
\begin{equation}      
C^Y_\ell =  C_{10}^Y \left (\frac{\ell}{10} \right )^{- \alpha_{Y}}   \hspace{1cm} Y = E,B,P,I_p,
\end{equation}
whose best fit parameters in the $2 \le \ell \le 25$ range are reported in
Table~\ref{power_spectra}. We perform a linear fit to the quantities $\log
C_\ell$ and $\log \ell$ giving greater weights to high order multipoles (since 
$C_\ell$ is an average of $(2\ell + 1) $ squared  harmonic amplitudes). In
addition, we compute the scalar spectrum $C^{I_p}_\ell$ of $I_p$ for comparison
with other works.

The two templates at 23~GHz are in good agreement in the overall intensity,
whereas the slopes are somewhat different, the 23--GHz--W $C^B_\ell$ being
flatter even if the slopes of the total polarization spectrum $C^P_\ell$ are
statistically equal. Differencies might be due to the different sky coverages of the two maps.
In fact, we have computed the APS for the 23--GHz--W template limited to the
coverage of 23--GHz--LF and have found a better agreement between the two
templates. This suggests the Southern sky contributes to make the overall slope
flatter.

A slight decrement of the slopes with increasing frequencies appeares in the
$C^B_\ell$ and $C^P_\ell$ spectra of our templates, the 90~GHz ones being the
flattest. However, these variations are within $2 \sigma$ of the slopes
themselves and a robust result cannot be claimed.

While the slopes of $C^E_\ell$ and $C^P_\ell$ are compatible with each other,
$C^B_\ell$ and $C^{I_p}_\ell$ are significantly flatter and this behaviour is
still not clear. In addition, $C^E_\ell$ and $C^P_\ell$ slopes are in agreement
with those previously calculated in real radio surveys (Tucci et al. 2000,
Baccigalupi et al. 2001, Bruscoli et al. 2002, Giardino et al. 2002). However,
one should take into account that a perfect agreement is not expected at all,
since our template and previous works differ in both the angular scale and the
frequency (1.4--2.7~GHz for the real surveys).

\section{High Galactic latitude emission}
\label{for_aps}

The templates presented here predict that the synchrotron polarized 
emission is concentrated on the Galactic Plane. As it already occurs for the
total intensity, this suggests us to search for the CMBP signal at high Galactic
latitude.

Let's us recall that this was not that obvious before the present work, since 
large and bright features exist, out of the Galactic Plane, in low 
frequency observations (see, for example, the Northern Galactic Spur in the 1.4~GHz
polarized data of Brouw \& Spoelstra 1976). 

\begin{figure}      
 \begin{center}  
 \includegraphics[width=1.0\hsize]{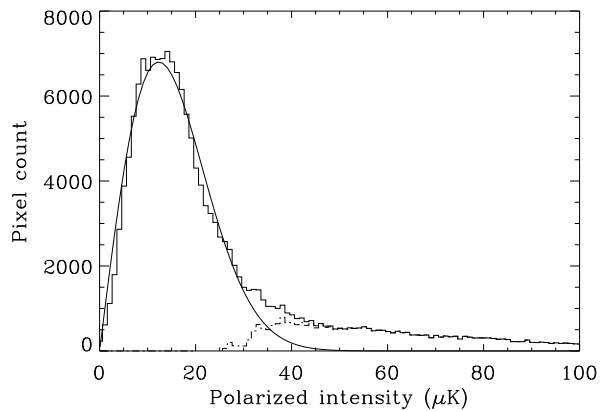}      
 \caption{Histogram of the polarized intensity distribution in the 23--GHz--W
 template. The solid curve represents the Rayleigh function used to fit the
 distribution e the dashed line represents the residual distribution.}     
 \label{pix_dis}     
 \end{center}  
 \end{figure}

To study the expected synchrotron emission at high Galactic latitudes we
mask off the Galactic Plane by using the  pixel--count histogram of 
the $I_p$ map at 23~GHz shown in Figure \ref{pix_dis}, in a way similar to that 
described by Bennett et al. (2003b). The intensity distribution is strongly
asymmetric, and pixels with the higher values correspond to the Galactic Plane
emission to be cut away. To set the value of the threshold defining
which pixels are to be discarded, we fit the peak of the
distribution and subtract the best--fit curve from the histogram. %
The peak  of the residual distribution was chosen by Bennett et al. (2003b)
as the threshold value. 

Differing from the Gaussian behaviour followed by temperature data
(Bennett et al. 2003b), the low emission part of the polarized intensity 
$I_p = \sqrt{Q^2 + U^2}$ is better represented by a Rayleigh distribution,
as it would be if $Q$ and $U$ were gaussian distributed:  
\begin{equation}
F(I_p) = A \,\, I_p \, e^{-{I_p^2}/{(2 \sigma^2)}}
\end{equation} 
where $A$ and $\sigma$ are parameters to be determined. 

\begin{figure}      
 \begin{center}  
 \includegraphics[width=0.6\hsize,angle=90]{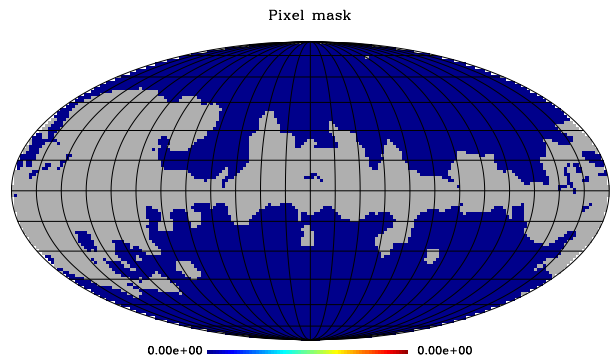}      
 \caption{The pixel mask in Mollweide projection. Grey pixels represent the 
 excluded region.}     
 \label{pix_mask}     
 \end{center}  
 \end{figure}

We fit the distribution in the $0$--$22~\mu$K range. The best--fit subtracted
histogram has a peak at 39.1~$\mu$K; however, we prefer to be more conservative
and set the threshold at 34.1~$\mu$K of the secondary peak. The pixel mask of
the retained pixels ($I_p < 34.1$~$\mu$K) covers about $68$\% of the sky and is
shown in Figure~\ref{pix_mask}.

\begin{figure*}      
 \begin{center}  
 \includegraphics[width=0.45\hsize]{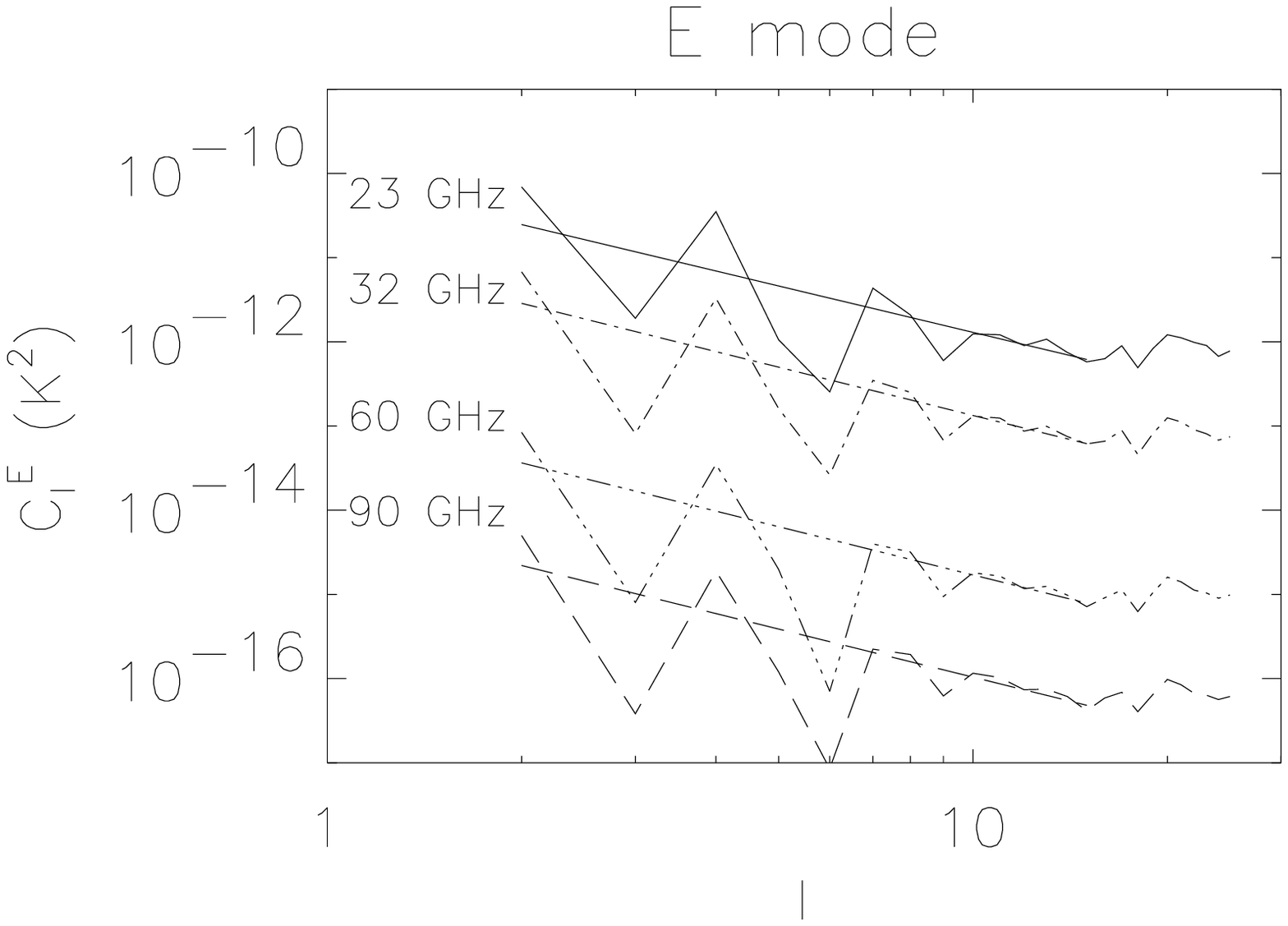}      
 \includegraphics[width=0.45\hsize]{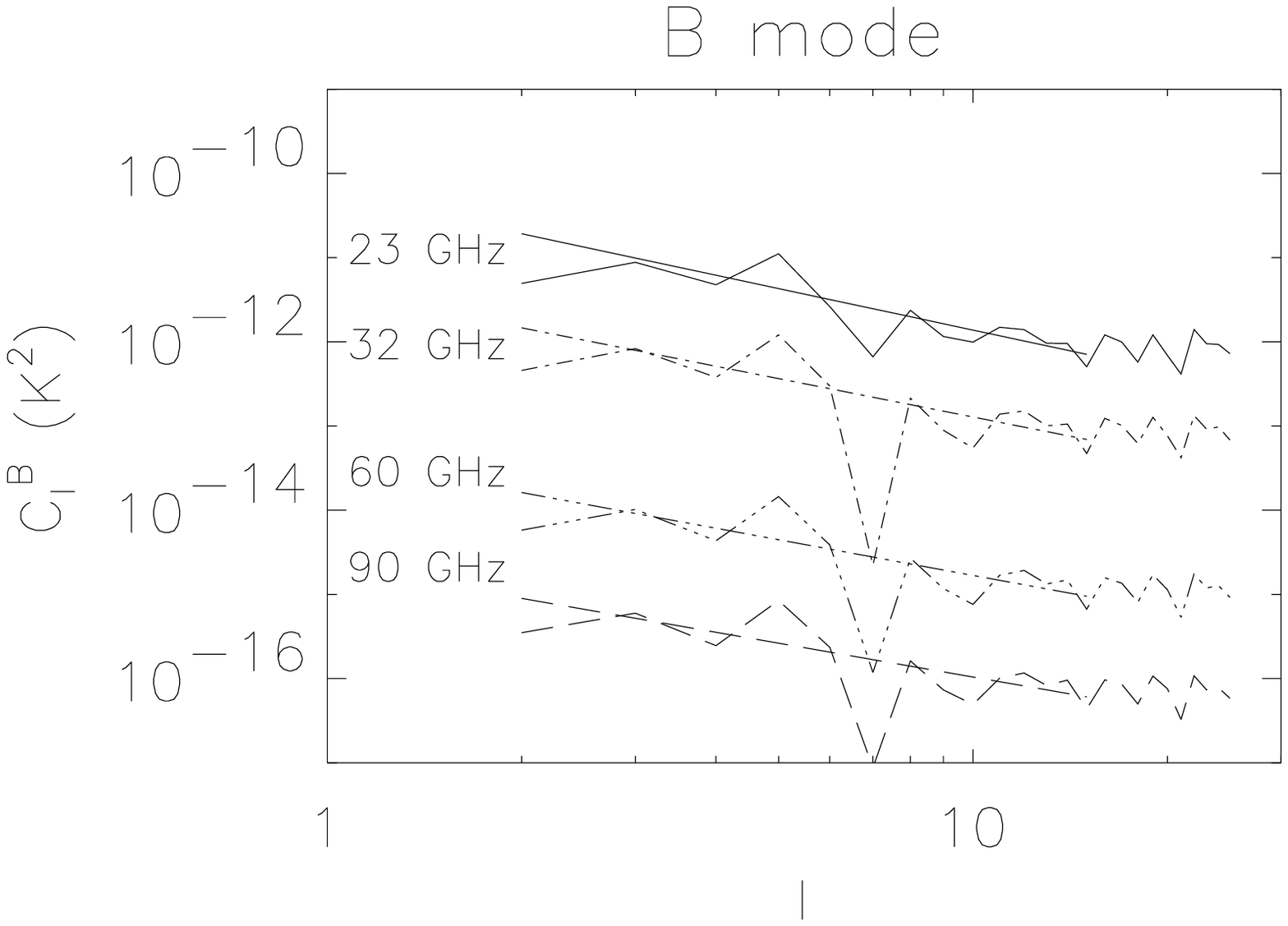}      
 \includegraphics[width=0.45\hsize]{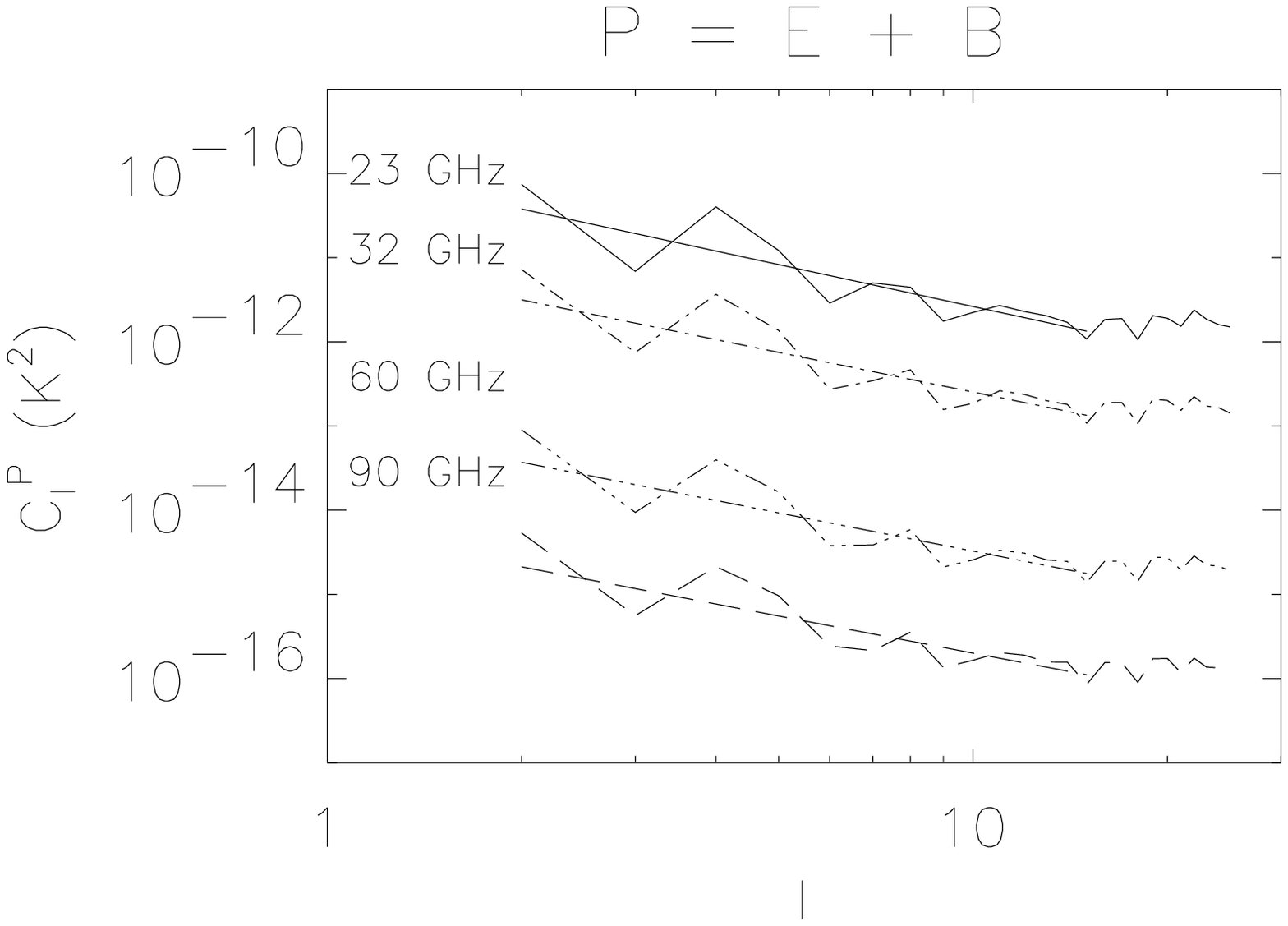}      
 \includegraphics[width=0.45\hsize]{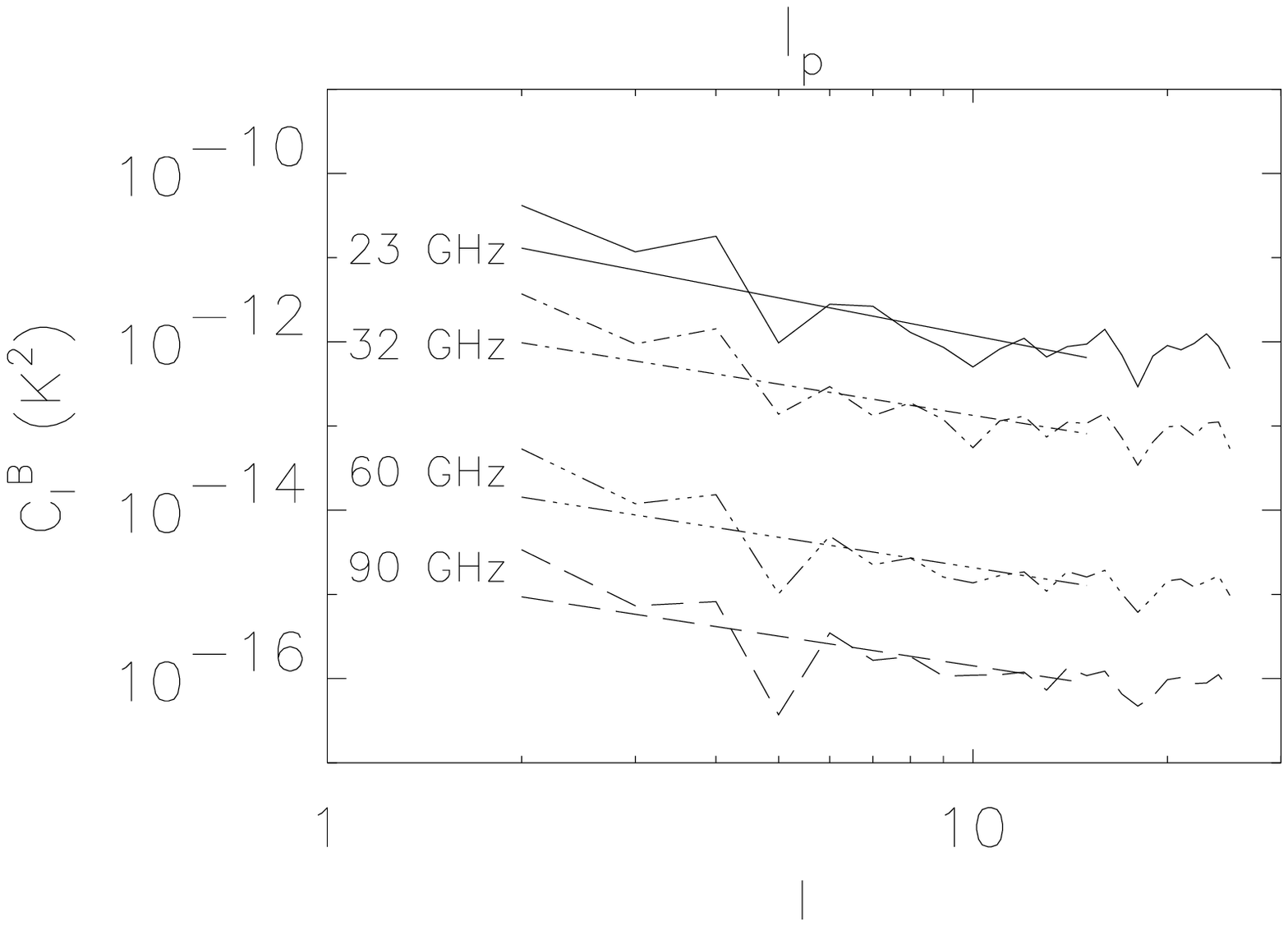}      
 \caption{$C^E_\ell$, $C^B_\ell$, $C^P_\ell$ and $C^{I_p}_\ell$ spectra 
 computed for the 23~GHz template with the high Galactic latitude mask applied 
 (dotted line). Spectra for the 
 32~(dashed), 60~(dashed--dotted) and
 90~GHz~(dashed--triple dotted) models 
 are shown as well.} 
 \label{eb_aps_masked}     
 \end{center}  
 \end{figure*}

APS computed on our template maps after applying this mask are shown in
Figure~\ref{eb_aps_masked}. Power spectra at high Galactic latitudes can
be represented by power laws apart from the region at large $\ell$, where they 
are observed to flatten, probably due to pixel noise. Therefore, we fit  the $2
< \ell < 15$ range and report the results in Table~\ref{power_spectra_masked}.
 \begin{table*}    
 \begin{center}    
 \centering
 \caption{Best fit APS slopes and amplitudes as obtained from our templates in
 the $2 \le \ell \le 15$ range after the pixel mask is applied.} 
 \begin{tabular}{|p{1cm}|*{8}{c|}|}    
 $\nu$ & $\alpha_E$ & $C^E_{10}$ & $\alpha_B$ & $C^B_{10}$ &
 $\alpha_P$ & $C^P_{10}$ & $\alpha_{I_p}$ & $C^{I_p}_{10}$\\
 (GHz) &  & $(\mu$K$^2)$ &  & $(\mu$K$^2)$ &  & $(\mu$K$^2)$ & & $(\mu$K$^2)$\\
\hline    
23        & $1.8 \pm 0.2$ & $1.3 \pm 0.1$      & $1.6 \pm 0.2$ & $1.4 \pm 0.1$ 
          & $1.7 \pm 0.2$ & $2.6 \pm 0.2$      & $1.5 \pm 0.2$ & $1.2 \pm 0.1$\\     
32        & $1.9 \pm 0.2$ & $0.13 \pm 0.01$    & $1.5 \pm 0.2$ & $0.13 \pm 0.01$ 
          & $1.6 \pm 0.2$ & $0.25 \pm 0.02$    & $1.2 \pm 0.2$ & $0.13 \pm 0.02$\\     
60        & $1.9 \pm 0.2$ & $(1.7 \pm 0.1) \times 10^{-3}$ & $1.4 \pm 0.2$ & $(1.7 \pm 0.1) \times 10^{-3}$ 
          & $1.5 \pm 0.2$ & $(3.3 \pm 0.2) \times 10^{-3}$ & $1.2 \pm 0.2$ & $(2.1 \pm 0.1) \times 10^{-3}$\\
90        & $1.9 \pm 0.2$ & $(1.04 \pm 0.07) \times 10^{-4}$    & $1.3 \pm 0.2$ & $(1.04 \pm 0.07) \times 10^{-4}$
          & $1.5 \pm 0.2$ & $(2.0 \pm 0.1) \times 10^{-4}$   & $1.2 \pm 0.2$ & $(1.4 \pm 0.1) \times 10^{-4}$
\label{power_spectra_masked}    
\end{tabular}    
\end{center}     
\end{table*}    

At each frequency, power spectra levels are about two orders of magnitude below
those including the Galactic Plane which, therefore, dominates the APS
calculated in Section \ref{tot_aps}.

Slopes are steeper than those computed on the overall templates, in particular
for $C^B_\ell$, suggesting that the flattening observed in the full--sky
templates is mainly due to Galactic Plane features. No significative flattening
with increasing frequency is found at high Galactic latitude.

\begin{figure*}      
 \begin{center}  
 \includegraphics[width=0.45\hsize]{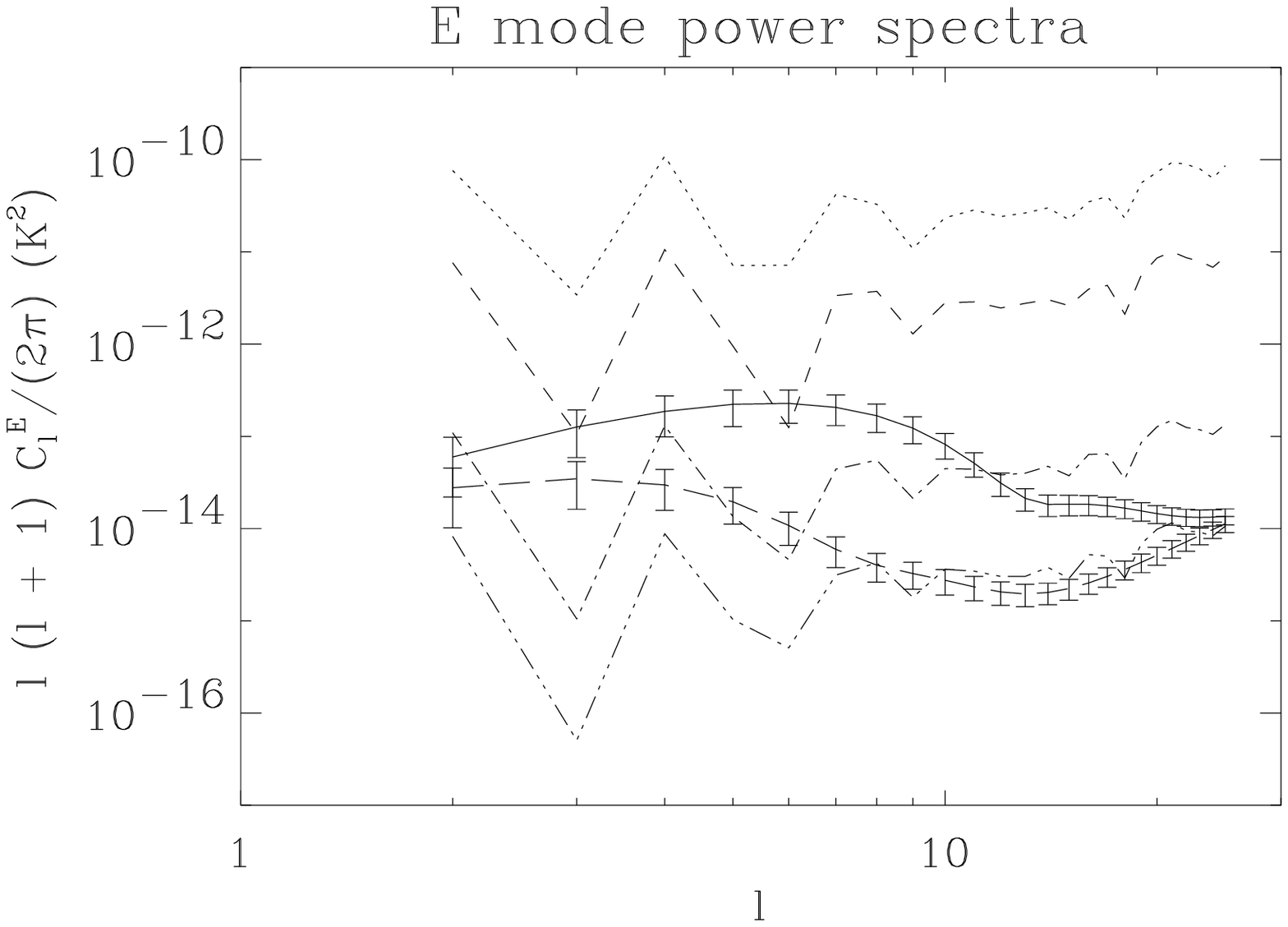}      
 \includegraphics[width=0.45\hsize]{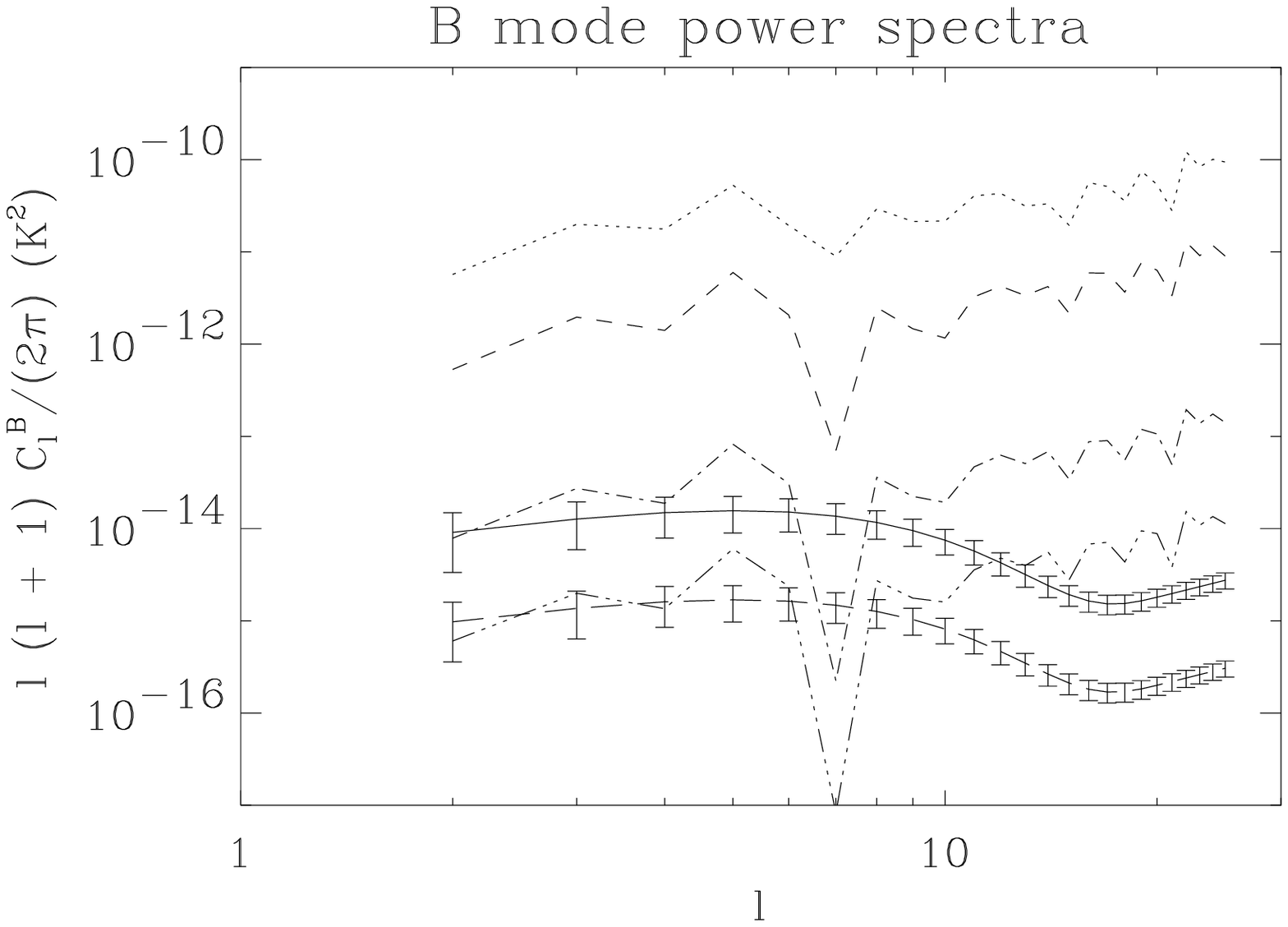}      
 \caption{Left: $C^E_\ell$ computed on the 23~GHz (dotted), 32~GHz (dashed),
 60~GHz (dashed--dotted) and 90~GHz models (dashed--triple dotted). The solid
 and long--dashed lines represent the CMB $E$--mode spectrum of the {\it
 concordance model} with $\tau = 0.17$ and $\tau = 0.05$, respectively. Erros
 bars represent the cosmic variance. Right: the same but for $C^B_\ell$. The
 solid and long--dashed lines represent CMB $B$--mode spectrum with $T/S = 0.1$
 and $T/S = 0.01$, respectively}
 \label{eb_mode}     
 \end{center}  
 \end{figure*}

Figure \ref{eb_mode} shows a comparison between our estimate of synchrotron
power spectra at high Galactic latitudes and expected CMBP angular spectra.
The power spectra of our templates are converted to CMB thermodynamic
temperature by multiplying them by the square of the conversion factor 
\begin{equation} \label{conversion_factor}
c  =\left( \frac{2\sinh \frac{x}{2}}{x} \right)^2
\end{equation}
where $x \equiv h \nu/kT_{cmb} \approx \nu/56.8$~GHz. CMBP angular spectra are
computed assuming cosmological--parameter values  as in the {\it concordance
model} determined by WMAP data (Spergel et al. 2003): $h_0 = 71$, $\Omega_b =
0.044$, $\Omega_{CDM} = 0.226$, $\Omega_{\Lambda} = 0.73$, $\tau = 0.17$ and
initial adiabatic fluctuations.

Our templates suggest that at 90~GHz CMBP $E$--mode measurements in the $2 <
\ell< 20$ multipole range should be only marginally contaminated by synchrotron
polarized radiation. Furthermore, even a less reionized Universe ($\tau = 0.05$)
should be accessible at this frequency. 

It is worth noticing that even relaxing our hypotheses of a steepening by 0.5 in
the spectral indeces between 23 and 90~GHz, e.g. simply adopting the
spectral--index map of Figure \ref{spectral_index}, the  normalization  of the
synchrotron power spectrum would be just a factor $\sim 4$ higher, which does
not substantially modify  the conclusions drawn above. In fact, the CMBP power
spectrum of models with $\tau = 0.17$ and $\tau = 0.05$ are about two and one
orders of magnitude larger than the synchrotron spectrum, respectively.

Obtaining a clean measurement of the CMBP $B$--mode looks more difficult. Large
values of the optical depth $\tau$ push a large--scale peak at $\ell < 10$ and
open a second window to detect this faint signal beyond that at $\ell \sim 100$
(Kamionkowski \& Kosowsky 1998). Figure~\ref{eb_mode} shows two models with the
same cosmological parameters as above, but with the addition of a gravitational
wave background with tensor to scalar fluctuation power ratio $T/S = 0.1$ and
$T/S = 0.01$, respectively. While the former looks reasonably free from
synchrotron contamination, at least in the low multipole range ($\ell < 10$),
the level of the latter is comparable with our estimate of the synchrotron
spectrum.

We can thus conclude that, for values $\tau = 0.17$ and $T/S > 0.1$, a CMBP
$B$--mode measurement on large angular scale free from synchrotron contamination
is achievable. For lower $T/S$ values, instead, detection of the $B$--mode would
be only possible in selected regions with low foreground emission, and large
enough to access the $\ell\sim 100$ peak. A good example is the patch observed
by the BOOMERanG experiment and selected for BaR-SPOrt (Cortiglioni et al.
2003), for which synchrotron contamination is expected to be under control
already for  $T/S > 0.01$ (Bernardi et al. 2003b). 

To complete our comparison between synchrotron and CMB, we compute the $P_{\rm
rms}$ of our masked template maps at the four frequencies (Table
\ref{prms_table}) using the equation (Zaldarriaga 1998)
\begin{equation}    
P_{\rm rms} = \sqrt{ \sum_{\ell} \frac {(2 \ell + 1)}{4 \pi} (C_{\ell}^E +
C_{\ell}^B)W_{\ell}} 
\end{equation}
where $W_{\ell}$ is the window function defined as 
\begin{equation}
W_{\ell} = e^{-\ell (\ell + 1) \sigma^2}
\end{equation}
and $\sigma = \mbox{FWHM}/\sqrt{8 \ln 2}$

At 90~GHz, the $P_{\rm rms}$ signal is already lower than the value computed in
B03 using the half faintest sky. Considering the 50\% faintest sky, a
$P_{\rm rms} \sim 0.057$~$\mu$K is obtained with the present model, showing that
the mean polarized value of the synchrotron background appears quite low in
large part of the sky.
\begin{table}    
\begin{center}    
\caption{$P_{\rm rms}$ of our templates at the four reference frequencies in
thermodynamic temperature. The $P_{\rm rms}$ is computed on the remaining 68\%
of the sky after applied the pixel mask. Bottom line reports the $P_{\rm rms}$
of the CMBP $E$--mode model with $\tau = 0.17$. The error on the CMBP $P_{\rm
rms}$ takes into account only the cosmic variance contribution.}
\begin{tabular}{|c|c|c|}    
$\nu$ (GHz) & $P_{\rm rms}$ ($\mu$K)\\     
\hline    
23 &  10\\     
32 &  3.2\\     
60 &  0.36\\     
90 &  0.095\\
\hline
CMBP $E$--mode ($\tau = 0.17$) & $0.49 \pm 0.03$\\
CMBP $E$--mode ($\tau = 0.05$) & $0.189 \pm 0.007$
\label{prms_table}    
\end{tabular}    
\end{center}     
\end{table}    

Table \ref{prms_table} shows that a $\tau = 0.17$ model has a $P_{\rm rms}$ for
the CMB $E$--mode which is about five times the $P_{\rm rms}$ of the high
latitude synchrotron foreground. This suggests that even an experiment having
just the sensitivity to detect the $P_{\rm rms}$ has good chances to measure the
optical depth $\tau$.

\section{Conclusions}
\label{conc}
In this work we generate template maps of polarized Galactic synchrotron
emission at 23, 32, 60 and 90~GHz using WMAP data.

We apply the method developed in B03 to the 23~GHz total intensity synchrotron
map released by the WMAP team. This greatly improves the B03 templates avoiding
the uncertainties in the extrapolation from low frequency. As for the previous
work, the angular resolution is limited to FWHM=$7^\circ$ mainly due to the
starlight polarization data sampling.

The basic template (23--GHz--W) is constructed at 23~GHz applying the method
directly to the WMAP data and it covers almost all the sky (94\%). Differing
from the 1.4~GHz B03 template where the Northern Galactic Spur represents an 
important feature, 23--GHz--W does not present relevant structures out of the
Galactic Plane. This replicates what already occurs in total intensity and
leaves most part of the sky with low synchrotron emission and useful for CMBP
investigations.

To achieve the templates at the other frequencies a spectral index map is
required to extrapolate the 23--GHz--W result. Differing from Bennett et al.
(2003b), which derived a spectral index map from 0.408 to 23~GHz with no
free--free subtraction in the 0.408~GHz map, we computed the frequency behaviour
using {\it pure} synchrotron maps at 1.4 and 23~GHz. The 23~GHz one is that
released by the WMAP team. The 1.4~GHz map has been achieved from low frequency
surveys by applying a Dodelson-like component separation. We use the Haslam et
al. 0.408~GHz and the Reich 1.4~GHz surveys for the Northern sky and the Haslam
et al. and the Jonas et al. 2.3~GHz ones for the Southern Hemisphere. The
results is the first almost all-sky spectral index map of the Galactic
synchrotron emission in the 1.4--23~GHz range. Furthermore, we add an extra 0.5
to the indeces to account for the steepening found by Bennett et al. (2003b) in
the 23--41~GHz range and this spectral index map is used to compute template
maps at 32, 60 and 90~GHz covering 92\% of the sky.

The spectral index map provides steeper indexes at high Galactic latitudes so
that the resulting templates have the Galactic Plane emission more and more
dominant from 23~GHz up to 90~GHz. This difference also explains the
disappearing of the Northern Galactic Spur in our templates. Unclear, instead,
it is the reason for such a flat frequency spectrum found in the Galactic Plane,
it being either intrinsic to the synchrotron emission (for instance due to a
young population of relativistic electrons) or due to a residual thermal
component.

Beyond the lack of relevant structures at high Galactic latitudes, the
differences among the templates of this work and the B03 one also concern the
APS. APS of our full--sky templates are dominated by the emission from the
Galactic Plane and flatter slopes, namely in the 1.4--1.6 range for $C^E_\ell$,
0.5--0.8 for $C^B_\ell$ and 1.2--1.5 for $C^P_\ell$. This general flattening is
still not clear.

A further unclear point is that the synchrotron APS slopes (in particular
$C^B_\ell$ and $C^P_\ell$) seem to flatten with increasing frequency. If
confirmed, these two aspects would confirm that low frequency APS cannot be
straightforwardly extrapolated from low to microwave frequencies.

The situation is less complex at high Galactic latitudes. In fact, limiting the
analysis out of the Galactic Plane, a steepening of the slopes occurs, which
become more similar to the B03 ones, and no significative dependency from the
frequency is found.

Interesting conclusions are drawn by the comparison of the high Galactic
latitude portion of our templates to CMBP as from the {\it concordance model} of
the WMAP first--year data. Indeed, the CMBP $E$--mode spectrum is about 2 orders
of magnitude above the synchrotron signal in our template at 90~GHz. Synchrotron
is thus unlikely to contaminate the cosmological signal. Furthermore, even a
less re-ionized Universe ($\tau = 0.05$) should be dominant over the Galactic
emission.

The situation looks different for the CMBP $B$--mode. Our template predicts that
this cosmological signal might be accessible at largest scales ($\ell < 10$)
only for a Tensor to Scalar fluctuation power ratio $T/S > 0.1$. To investigate 
models with lower values of $T/S$ one should restrict to selected low
synchrotron emission areas large enough to detect the $\ell \sim 100$ peak. An
example is the sky patch observed by the BOOMERanG experiment where models with
$T/S > 0.01$ seem to be accessible (Bernardi et al. 2003b).

Our templates obviously rely on the assumption that WMAP's dust-correlated
signal at 22~GHz is genuine synchrotron, as supported by Bennet et al. (2003b).
However several authors (de Oliveira--Costa et al. 2003, Banday et al. 2003,
Lagache 2003) recently support the opposite view, that WMAP's data may be more
consistent with anomalous dust emission. For instance, while the  spinning-dust
model originally proposed by Draine \& Lazarian (1998) may meet some difficulty,
emission by very small grains is supported by Lagache (2003).

Considering that dust polarization level is likely less than 5\%
rather than the 15-30\% of the synchrotron, a measurement of polarization at
22-32~GHz could help to decide on this issue: should the polarization level be
found a factor $\sim 5$ below the predictions of our template, the role of
anomalous dust emission would be supported. 

Also, in such a case, the contamination of CMBP maps at 90~GHz would be even 
smaller than considered in our discussion above, and prospects for the detection
of a cosmological B mode would be slighly better.

\section*{Acknowledgments}

This work has been carried out in the frame of the SPOrt experiment, a programme
funded by ASI. G.B. acknowledges a Ph.D. ASI grant. We thank an anonymous
referee for useful comments. We acknowledge the use of CMBFAST package. Some of
the results in this paper have been derived using the
HEALPix\footnote{http://www.eso.org/science/healpix/} (Gorski, Hivon, and
Wandelt 1999) package. We acknowledge the use of the
data\footnote{http://lambda.gsfc.nasa.gov/} made publicy available by the WMAP
team.

\bsp     
     
\label{lastpage}     
     
\end{document}